\documentclass[twocolumn]{aastex701}
\pdfoutput=1

\usepackage{amsmath}

\newcommand{\logg} {\log \textsl{\textrm{g}}}

\newcommand{\Te} {T_{\rm eff}}

\newcommand{\bra} {\langle}
\newcommand{\ket} {\rangle}

\begin{document}

\title{Improved Stark Broadened Profiles for Neutral Helium Lines Using Computer Simulations}

\author{Patrick Tremblay} 
\affiliation{D\'epartement de Physique, Universit\'e de Montr\'eal, C.P.~6128, Succ.~Centre-Ville, Montr\'eal, Qu\'ebec H3C 3J7, Canada}
\email[show]{patrick@astro.umontreal.ca}

\author{Alain Beauchamp} 
\affiliation{D\'epartement de Physique, Universit\'e de Montr\'eal, C.P.~6128, Succ.~Centre-Ville, Montr\'eal, Qu\'ebec H3C 3J7, Canada}
\email{alainbeauchamp1905@outlook.com}

\author{Pierre Bergeron} 
\affiliation{D\'epartement de Physique, Universit\'e de Montr\'eal, C.P.~6128, Succ.~Centre-Ville, Montr\'eal, Qu\'ebec H3C 3J7, Canada}
\email[show]{bergeron@astro.umontreal.ca}

\author{Antoine Bédard} 
\affiliation{Department of Physics, University of Warwick, Coventry CV4 7AL, UK}
\email{Antoine.Bedard@warwick.ac.uk}

\shortauthors{Tremblay et al.}
\shorttitle{Stark Profiles for He {\sc i} Lines}

\begin{abstract}
The study of Stark broadening of neutral helium lines, despite
significant advances over recent decades, has not led to updated large
grids of helium line profiles relevant to the spectroscopic study of
helium-rich stars. While the semi-analytical approach based on the
standard Stark broadening theory is efficient for generating such
grids, it presents challenges in incorporating additional physical
effects into the model. Motivated by recent studies that highlight
potential issues with line profiles in the context of white dwarf
stars, this paper leverages advances in computer simulations to create
a new grid of line profiles for 13 neutral helium lines in the optical
range. These profiles cover densities ranging from $10^{14}$ to
$6\times10^{17}$ cm$^{-3}$ and temperatures from 10,000 K to 40,000 K,
with the exception of the narrower He {\sc i} $\lambda 4713$ line, for
which the profile grid begins at $10^{15.5}$ cm$^{-3}$. The primary
goal of this research is to present the new grid and compare it with
both the semi-analytical approach and other simulation results. By
doing so, corrections to the previous grid will be explored, providing
a foundation for future studies that utilize this updated grid.  We
also examine the impact of these new profiles on the
determination of physical parameters for a range of astrophysical
objects, including DB white dwarfs and other helium-rich stars.
\end{abstract}


\section{Introduction} \label{intro}

Recent studies on white dwarf stars
\citep{Bergeron2011,Genest2019b,Bergeron2019,Cukanovaite2021} have
revealed challenges in determining physical parameters using
traditional methods, specifically the photometric and spectroscopic
techniques (see, e.g., \citealt{Bergeron1997} and
\citealt{Bergeron92}, respectively). The photometric technique relies
on fitting the overall spectral energy distribution, while the
spectroscopic technique focuses on fitting individual line profiles,
making it more sensitive to atomic physics details. In an effort to
reconcile the physical parameters derived from both methods, several
lines of investigation have been proposed, including a thorough review
of the line profile grids used in spectroscopic analyses. For
helium-line DB white dwarfs, these grids originate from the work of
\citet[][hereafter B97]{Beauchamp97}. Although this grid has been
dependable over the years, a timely update is essential to align with
modern calculations. This effort began with the study by
\cite{PTremblay2020}, which utilized computer simulations. Our goal is
to advance this initiative by providing a major update to the
simulation-based line profiles, incorporating improvements suggested
by \cite{Rosato2020} and \cite{Cho2022}.

The computer simulation method for generating line profiles is not
new; it has existed since the time when semi-analytical methods based
on the standard Stark broadening theory were the preferred option
(B97). However, it has consistently been constrained by its
computational and time-intensive demands. The original motivations
behind using simulations were twofold: to go beyond the impact
approximation for electrons, and to model the effects of ion dynamics
on line profiles, a factor often neglected in most studies relying on
semi-analytical profiles (with the exception of \citealt{Barnard74}
and \citealt{Barnard75}, who generated exhaustive line profile grids
for He {\sc i} $\lambda\lambda$4471 and 4922).

The foundational work on the computer simulation method was introduced
by \cite{Stamm79}, \cite{Stamm84}, and later refined by
\cite{Hegerfeldt88}.  It was not until the advancements introduced by
\citet{Gigosos85} and \citet{Gigosos87} that the simulation approach
gained significant attention. Following these developments, many
efforts were made to improve the balance between computational time
and the quality of the line profiles generated through simulations
\citep{Calisti88, Frerich89, Alexiou99b, Alexiou99}. This progress was
particularly evident in studies focused on the hydrogen atom
\citep{Gigosos96, Halenka96, Sorge2000, Olchawa2002, Wujec2002,
  Halenka2002, Gigosos2003, Olchawa2004, Ferri2014, GomezPhD} and the
helium atom \citep{Gigosos2009, Lara2012, PTremblay2020} as emitters.

All of these efforts used the autocorrelation function approach to
generate line profiles. However, some lines showed significant noise
in the final results \citep{PTremblay2020}, especially in the far
wings, a finding later confirmed mathematically by \cite{Rosato2020}
and demonstrated by \cite{Cho2022} for hydrogen lines. These key
studies highlight the importance of directly applying the power
spectrum equation for line profile generation, avoiding the use of the
autocorrelation function. This shift enhances the time-quality ratio
of the method.

We begin in Section~\ref{sec:THEO} by outlining the theoretical
framework of the standard Stark broadening theory, along with the
simulation methodology and its implementation details, including the
power spectrum method and the particle-generation
procedure. Section~\ref{sec:comp0} then concentrates on the extraction
of He {\sc i} line profiles and presents intermediate validation
results. The subsequent analysis offers a detailed comparison between
the simulated line profiles and previously published calculations. In
Section~\ref{sec:appl}, we examine the potential impact of these new
profiles on the determination of physical parameters for a range of
astrophysical objects, including DB white dwarfs and other helium-rich
stars. Concluding remarks are finally presented in
Section~\ref{sec:conc}.

\section{Theoretical Background \label{sec:THEO}}

\subsection{Standard Stark Broadening Theory}

Following the formulation introduced by \citet{Anderson1949}, the
spectral line shape can be expressed as the real part of the Fourier
transform of the autocorrelation function $C(t)$:
\begin{equation}
\label{eq:stark9}
I(\omega)=\frac{1}{\pi}\mathrm{Re}\int_0^\infty dt\,e^{i\omega t} C(t),
\end{equation}

\noindent
where the autocorrelation function is given by
\begin{equation}
\label{eq:C68}
C(t) = \sum_{ab}
\rho_a
\langle a|\mathbf{d}|b\rangle\
\cdot
\Big\{
\langle a| \ U^\dagger(t,0) \mathbf{d}
U(t,0)|b\rangle \Big\}_{\mathrm{av}}.
\end{equation}

\noindent
Here, the states $|a\rangle$ and $|b\rangle$ correspond to the upper
and lower levels of the radiative transition of interest, $\rho_a$
denotes the statistical weight of state $a$, and $\mathbf{d}$ is the
dipole operator in the Schrödinger picture. The operator $U(t,0)$
represents the time evolution of the emitter under the influence of
external perturbations and is governed by the total Hamiltonian
\begin{equation}
H = H_0 + V(t),
\label{eqn:H_HeI}
\end{equation}

\noindent
where $H_0$ is the unperturbed atomic Hamiltonian and $V(t)$ describes
the time-dependent interaction with the surrounding plasma
perturbers. The expression for $C(t)$ includes a thermal average over
all dynamical configurations of the perturbers, which generally
comprise both electrons and ions.

In the standard Stark broadening theory, the large disparity between
the characteristic velocities of electrons and ions is exploited by
adopting the quasi-static approximation for the slowly moving ions and
the impact approximation for the rapidly moving electrons
\citep{Baranger58a,Baranger62,Smith69B,Barnard69,Sahal69a,Sahal69b,Griem74,Gigosos2014}.

By neglecting interactions between ions and electrons, the thermal
average can be factorized into independent averages over ionic and
electronic configurations. As a result, the average over static ion
configurations can be reformulated as an average over ion-induced
electric fields, weighted by the microfield distribution function
$W_R(F)$. This distribution accounts for spatial correlations among
ions arising from their mutual Coulomb interactions and is
characterized by the parameter $R$, defined as the ratio of the mean
interionic distance to the Debye radius \citep{Hooper68}.

Within this framework, the autocorrelation function governing the
spectral line profile becomes
\begin{equation}
\label{eqn:CtStandard}
C(t) = \int_0^\infty dF \ W_R(F)\  {C_e(t;F)}_{{\rm av}_e},
\end{equation}

\noindent
where $C_e(t;F)$ denotes the electron-induced autocorrelation function
in the presence of a static electric field $F$. The function
$C_e(t;F)$ retains the same formal structure as
Equation~(\ref{eq:C68}), with the thermal average now restricted to
the ensemble of electron trajectories. The corresponding time
evolution is governed by the Hamiltonian
\begin{equation}
\label{eq:Ht}
H(t) = H_0 + eFZ + V_e(t),
\end{equation}

\noindent
which includes a constant electric field $F$ aligned along the
$z$-axis, $Z$ the $z$-component of the position operator of the
emitting electron, and the time-dependent interaction $V_e(t)$ between
the emitter and the electrons.

The evolution operator associated with Equation~(\ref{eq:Ht}) is
evaluated within the framework of the impact approximation. This
approximation relies on three key assumptions: (1) strong collisions
are temporally isolated and do not overlap, (2) the cumulative effect
of weak, overlapping collisions can be adequately described by
truncating the Dyson series at second order, and (3) the
characteristic collision duration is much shorter than the inverse of
the spectral line half width at half maximum (the completed collision
assumption).

After performing the thermal average over electron trajectories and
applying the inverse Fourier transform defined in
Equation~(\ref{eq:stark9}), the resulting spectral line profile is
obtained as
\begin{equation}
\label{eq:QS2}
I(\omega)  = \int_0^\infty dF \ W_R(F) \ I_e(\omega;F),
\end{equation}

\noindent
which represents a weighted superposition of electron-broadened line
profiles corresponding to fixed values of the ionic microfield $F$.

Each component profile $I_e(\omega;F)$ is given by
\begin{equation}
\label{eq:Ie}
\begin{split}
I_e(\omega;F) &= -\frac{1}{\pi}
\sum_{\alpha \alpha' \beta \beta'}
\langle \alpha |{\bf d}|\beta\rangle
\langle \beta'|{\bf d}
|\alpha'\rangle\\
&\quad\langle \alpha | \langle \beta |
\left[i(\omega-\omega_{\alpha \beta})+\Phi\right]^{-1}
|\alpha' \rangle |\beta'\rangle,
\end{split}
\end{equation}

\noindent
where $\omega_{\alpha \beta}$ denotes the transition frequency between
the perturbed upper and lower states $|\alpha\rangle$ and
$|\beta\rangle$, respectively. The operator $\Phi$ encapsulates the
electron-induced broadening effects within the impact
approximation. For clarity, the dependence of the perturbed states,
energy levels, and the operator $\Phi$ on the electric field $F$ is
not explicitly indicated. It is further assumed in this work that all upper
states share the same statistical weight, which can therefore be
omitted, as it cancels out when the final profile is normalized to
unity.

Within the impact approximation, the collision operator $\Phi$ is
obtained by integrating over the full trajectory of a single electron
and subsequently averaging over all possible trajectories. Under the
classical path approximation, the electron is assumed to follow a
straight-line trajectory at constant velocity. The averaging procedure
therefore involves integration over all such trajectories,
parameterized by three angular variables, the electron velocity $v$,
the impact parameter $\rho$, and the time of closest approach. The
integral over $\rho$ exhibits formal divergences at both small and
large values, which are treated by introducing appropriate cut-offs to
ensure convergence in practical calculations.

In the far wings of the spectral line, the assumptions underlying the
completed collision approximation of impact theory break down. In this
regime, the one-electron theory developed by \citet{Baranger62}
provides a more appropriate description. This approach accounts for
the contribution of isolated, non-overlapping strong collisions and
describes the gradual transition of the line shape toward a
quasi-static profile. Within the one-perturber framework, the
definition of the collision operator $\Phi$ differs from that employed
in the standard impact theory. Notably, no rigorous theoretical
prescription exists for smoothly connecting the two regimes. A
practical implementation of this hybrid approach is described in B97.

Over the past three decades, spectroscopic analyses of DB white dwarfs
have relied on the He {\sc i} line profile tables of B97. These tables
have also been employed in more recent studies of other classes of
stars \citep{Przybilla2021,Jeffery2024,Latour2025}. The profiles were
computed using the standard Stark broadening theory as implemented by
\citet{Barnard69} and \citet{Barnard70}, with the inclusion of the
one-electron theory of \citet{Baranger62}. Line dissolution was
incorporated following the prescription of \citet{Seaton90}, by
restricting the integral in Equation~(\ref{eq:QS2}) to electric field
strengths below the critical value above which the upper levels of the
transition are dissolved.

The limitations of the standard theory outlined above underscore the
need to update existing grids of line profiles. In the context of DB
white dwarfs, the far wings of the line profiles contribute to the
pseudo-continuum and can influence the atmospheric parameters derived
using the spectroscopic method. The treatment of the transition between
the impact and one-electron regimes therefore introduces a source of
uncertainty in this region. Moreover, for the spectral analysis of
lower-density objects, such as main-sequence and helium-rich stars,
ion dynamics can significantly affect the line core and, when present,
the forbidden components. More generally, conclusions drawn from
analyses based on the B97 profile tables should be reassessed using
more modern calculations.

The objective of the simulation method presented in this work is to
eliminate most of the approximations inherent in the standard Stark
broadening theory while preserving the essential features of the
semi-classical framework (see below). As such, this method is well suited to
generate a new generation of Stark-broadened He {\sc i} line profiles.

\subsection{Computer Simulations}

Computer simulations currently represent the state of the art for the
calculation of Stark-broadened line profiles. This approach has been
successfully applied to numerous hydrogen lines
\citep{Gigosos96,Cho2022} as well as to neutral helium lines
\citep{Gigosos2009,Lara2012,Gigosos2014b,PTremblay2020}. Compared to
the standard Stark broadening theory, computer simulations mitigate
several inherent approximations, at the expense of a substantially
increased computational cost. Nevertheless, the method remains
sufficiently efficient to allow the generation of extensive grids of
line profiles covering the wide range of thermodynamic conditions
encountered in stellar atmospheres.

Unlike the standard theory, the simulation method correctly handles
the superposition of collisions regardless of their duration and is
not restricted to a perturbative treatment. In particular, the
transition between the two electron-broadening regimes—namely, the
impact regime dominating the line core and the one-electron regime
relevant in the wings—as well as ion dynamics, are treated naturally
within a unified framework that explicitly follows the time evolution
of all perturbers. At the same time, computer simulations retain
several advantages of the standard theory, most notably the ability to
model forbidden components, which play a crucial role in the formation
of helium line profiles. Indeed, many forbidden components are
observed in the spectra of DB white dwarfs
\citep{Liebert1976,Beauchamp1995} and other helium-rich stars
\citep{Jeffery1998,Przybilla2021}.

The simulation method relies on the validity of the semi-classical
approximation, in which the emitting atom is treated as a quantum
subsystem immersed in a time-varying electric field generated by
well-localized charged perturbers. These perturbers follow classical
trajectories determined by their respective charges and by the charge
state of the emitter. When the emitter is neutral, the perturbers move
along straight-line trajectories at constant velocity.

The line profile associated with a given radiative transition is
obtained as an arithmetic average over a large ensemble of dynamic
perturber configurations. Each configuration produces a temporal
sequence of electric fields that perturb the emitting atom. For a
given temporal sequence, the Hamiltonian governing the evolution of
the emitter is still given by Equation~(\ref{eqn:H_HeI}). In the
present work, only the first nonvanishing term of the multipole
expansion of the interaction $V(t)$ is retained, namely the dipole
term:
\begin{equation}
V(t) = e{\bf F}(t) \cdot {\bf R}
\end{equation}

\noindent
where ${\bf F}(t)$ denotes the electric field at the position of the
emitter and ${\bf R}$ is the position operator of the radiating
electron. Higher-order terms in the multipole expansion have been
considered by some authors in the case of hydrogen lines
\citep{Cho2022}.

The analysis of the electric field fluctuations experienced by the
emitter is restricted to contributions from charged particles located
within a spherical region, referred to as the simulation volume, whose
radius is of the order of the Debye length. The generation of a
temporal sequence of electric fields involves two distinct steps: the
initialization of a distribution of charged perturbers and the
reinjection of new perturbers as the simulation proceeds. During
initialization, perturbers of the relevant species are placed within
the simulation volume with initial position and velocity vectors drawn
from a random number generator in a manner consistent with Boltzmann
statistics. As time evolves, particles initially present within the
simulation volume may exit the sphere and are replaced by new
particles of the same type. This reinjection procedure ensures that
the joint statistical distribution of particle positions and
velocities remains stationary over time. Throughout the simulation,
the emitting atom is fixed at the center of the sphere.

The assumption of negligible interactions among perturbers implies
straight-line trajectories within the simulation volume, which does
not fully reflect the physical conditions of a plasma. To account for
inter-particle interactions at least at a first-order level, a
Debye-screened Coulomb potential is employed to describe the
electrostatic interaction between the perturbers and the emitter
\citep{Gigosos2003,Stambulchik2007}, with the Debye length computed
under the assumption of electron and ion screening. The statistical
distribution of the magnitude of the electric fields generated at the
center of the simulation volume, obtained from the ensemble of
perturber configurations, is found to be consistent with the expected
microfield distribution for a neutral emitter under the corresponding
thermodynamic plasma conditions \citep{Gigosos96,Cho2022}. Moreover,
this potential ensures that the electric field produced by an
individual perturber in the immediate vicinity of the emitter remains
Coulombian, as physically required.

Within the simulation framework, the thermal average over all
perturbers appearing in Equation~(\ref{eq:C68}) is approximated by an
average over a large but finite number $N$ of dynamic
configurations. The autocorrelation function $C(t)$ is therefore
computed as the mean of $N$ individual autocorrelation functions
$C_i(t)$, each corresponding to a specific configuration $i$. As time
increases, the dipole moment of the emitter becomes progressively
decorrelated from its initial value, causing the contribution of the
integral to diminish and eventually approach zero; at this point,
$C(t)$ can be considered converged. An analysis of the convergence
behavior of $C(t)$ thus provides a practical criterion for defining
the upper integration limit in Equation~(\ref{eq:stark9}),
corresponding to the total duration of the simulation.

Alternatively, the line profile can be computed using the power
spectrum method. As shown by \citet{Rosato2020} and \citet{Cho2022},
this approach yields a significantly improved signal-to-noise ratio
compared to the direct autocorrelation method. For this reason, the
power spectrum method is adopted in the present work to compute the
line profiles. Nevertheless, the autocorrelation function remains
useful for determining the total simulation duration for a given
simulation volume, that is, the number of time steps required for a
chosen time step $\Delta_t$.

\subsection{The Power Spectrum Method}

Using the same notation as in Equation~(\ref{eq:C68}), the power
spectrum associated with a given line transition is defined as
\begin{equation}
P(\omega) \propto \omega^4
\lim_{T\to +\infty} {1\over T} \sum_{ab} \rho_a
\Big< \Big| \int_{-T/2}^{T/2} dt \bra b | {\bf d}(t)|a \ket e^{i\omega t} \Big|^2 \ \big>_{\rm av},
\end{equation}

\noindent
where ${\bf d}(t)$ is the dipole operator of the emitting atom in the
Heisenberg picture \citep{Baranger58b,Baranger62,Rosato2020}, and $T$
denotes the time window over which the signal is observed or
computed. The slowly varying prefactor $\omega^4$ is omitted in the
detailed calculations, yielding the line profile
\begin{equation}
I(\omega) \propto
\lim_{T\to +\infty} {1\over T} \sum_{ab} \rho_a
\Big< \Big| \int_{-T/2}^{T/2} dt \bra b | {\bf d}(t)|a \ket e^{i\omega t} \Big|^2\ \Big>_{\rm av},
\label{eqn:Lomega_HeI}
\end{equation}

\noindent
which is subsequently normalized to unit area. Following the reasoning
previously discussed for the autocorrelation function, the line
profile is computed as the average of $N$ individual realizations,
\begin{equation}
I(\omega) = \frac{1}{N} \sum_{i = 1}^N I_i(\omega),
\label{eqn:Lomega2}
\end{equation}

\noindent
where $I_i(\omega)$ is obtained from the $i^{\rm th}$ temporal
sequence of dipole moments.

The power spectrum method is formally defined through the limiting
process $T \rightarrow +\infty$ (Equation~\ref{eqn:Lomega_HeI}). In
practice, however, this formulation does not provide guidance on the
minimum simulation duration required to obtain a profile free from
numerical artifacts or systematic biases arising from the finite time
window. An exploratory phase is therefore required to determine an
appropriate value of $T$ for each spectral line and set of
thermodynamic conditions considered in this work.

Numerical tests performed on the He {\sc i} $\lambda4471$ line show
that the normalized profile intensity depends sensitively on the
adopted value of $T$, particularly in the line wings at low electron
density. For insufficient simulation durations, the wings are
systematically overestimated, while the profile progressively
converges toward its asymptotic shape as $T$ increases. The formal
limiting process is therefore unnecessary, provided that the
simulation duration is chosen to be sufficiently long. This
requirement must be assessed separately for each spectral line and set
of plasma conditions (see Section~\ref{subsec:ID} for details).

Because the reference time of the evolution operator is arbitrary in
the Heisenberg representation, the final expression for the line
profile corresponding to the $i^{\rm th}$ configuration becomes
\begin{equation}
I_i(\omega) \propto
\sum_{ab} \rho_a \Big| \int_{0}^{T} d\tau \bra b | {\bf d}(t)|a \ket e^{i\omega \tau} \Big|^2,
\label{eqn:Liomega2}
\end{equation}

\noindent
which remains fully compatible with the numerical simulation framework.

\subsection{Generation and Reinjection of Particles}

As described in previous studies
\citep{Gigosos96,PTremblay2020,Cho2022}, the simulation approach
combines an initialization phase with a particle reinjection
process. Perturbing particles—either charged ions or electrons—are
initially placed inside a spherical simulation volume of radius $R$ at
$t=0$. As the simulation evolves, particles that leave the volume are
removed and replaced by new particles of the same species that enter
the volume, while the emitting atom remains fixed at the center of the
simulation sphere. Assuming a neutral emitter, the perturbers follow
straight-line trajectories within the sphere,
\begin{equation}
\label{eq:distradimp2article}
{\bf r}(t)=b(\hat{v_1}\cos\alpha+\hat{v_2}\sin\alpha)+\hat{v_3}v(t-t_0),
\end{equation}

\noindent
where $v$ is the constant particle velocity, $\hat{v_3}$ is the unit
vector along the velocity direction, $t_0$ is the time of closest
approach, $(\hat{v_1},\ \hat{v_2},\ \hat{v_3})$ form an orthonormal
basis, and the angle $\alpha$ specifies the orientation of the
trajectory in the plane defined by the velocity vector and the point
of closest approach of magnitude $b$. Once a particle exits the
simulation volume, it is removed from the simulation and does not
re-enter.

The particle generation procedure is designed to ensure that, within a
given simulation volume, the statistical properties of the system are
preserved at all times and across all simulations running in
parallel. In particular, the random generation of the angle $\alpha$
and of the orthonormal basis ensures isotropy and homogeneity during
both the initialization phase and the reinjection process.

At initialization, the joint probability distribution of $t_0$, $b$,
and $v$ is given by
\begin{equation}
\label{eq:heger4article}
p(t_0, b, v) \propto bv^3e^{-v^2/v_T^2},
\end{equation}

\noindent
which can be expressed as the product of conditional distributions,
\begin{equation}
\label{eq:heger5article}
p(t_0, b, v) = p(t_0|b, v)p(b|v)p(v),
\end{equation}

\noindent
subject to the additional constraint that the perturbers must lie
within the simulation volume,
\begin{equation}
\label{eq:gogo3}
|t_0| \le t_{\rm max}\ \equiv \frac{\sqrt{R^2-b^2}}{v}.
\end{equation}

\noindent
The joint distribution and the above constraint imply that
\begin{equation}
\label{eq:gogo5}
p(t_0|b, v)={1\over 2t_{\rm max}}= {v\over 2\sqrt{R^2-b^2}},
\end{equation}

\noindent
which corresponds to a uniform distribution over the allowed range of
$t_0$, and that
\begin{equation}
\label{eq:gogo6}
p(b|v) ={3\over R^3} b \sqrt{R^2-b^2}\ \equiv p(b),
\end{equation}

\noindent
which is independent of $v$, while $v$ itself follows the Maxwell
velocity distribution. Consequently, the variables $v$ and $b$ can be
generated independently for each particle using random number
generators. Once $v$ and $b$ are drawn, the corresponding value of
$t_{\rm max}$ is computed and $t_0$ is sampled from a uniform
distribution over the interval defined by Equation~(\ref{eq:gogo3}).

Since our previous work \citep{PTremblay2020}, several modifications
have been introduced to the particle reinjection procedure to improve
its physical realism. The method described by Tremblay et al.\ was
based on the approach originally proposed by \citet{Gigosos96}. Its
main conceptual limitation is that it enforces fluctuation-free
distributions during the simulation: the Maxwell velocity distribution
and the corresponding distribution of the distance of closest approach
$b$ are satisfied exactly at all times.

A more physically faithful approach, recently demonstrated by
\citet{Cho2022}, is adopted here. The key idea is that the statistical
distributions of the velocity $v$ and impact parameter $b$ for
incoming particles must match those of the particles that leave the
simulation volume. Particles with longer sphere-crossing times
($2t_{\rm max}$) exit the volume less frequently. Consequently, the
joint distribution $p_{\rm in}(v,b)$ of incoming particles must be
proportional to the distribution of particles within the simulation
volume at initialization, divided by the crossing time. This reasoning
leads to
\begin{equation}
p_{\rm in}(b|v) = {2\over R^2}b \equiv p_{\rm in}(b),
\end{equation}

\noindent
and
\begin{equation}
p_{\rm in}(v) = {2\over v_T^4} v^3 e^{-{v^2/v_T^2}},
\end{equation}

\noindent
after proper normalization.

With this reinjection scheme, the sampled distributions of $v$ and $b$
at any given time are allowed to deviate slightly from their
theoretical forms due to statistical fluctuations. This behavior is
physically more realistic, as the distributions are respected on
average rather than enforced deterministically. In addition, this
approach naturally facilitates extensions to more general particle
trajectories in future developments.

A further refinement of the reinjection process consists in allowing
particles to be reinjected within a thin spherical shell, rather than
strictly at the boundary of the simulation volume. The reinjected
particle is placed within a shell of variable thickness determined by
its velocity. The parameter $t_0$ associated with a reinjected
particle is then given by
\begin{equation}
\label{eq:lagvalue}
t_0 = \frac{\sqrt{(R - \Delta r)^2 - b^2}}{v},
\end{equation}

\noindent
where $\Delta r$ is a uniformly distributed random variable in the
interval $[0,\,v\Delta_t]$, representing the maximum distance the
particle could have traveled during the time step $\Delta_t$. This
correction prevents the reinjection of a new particle from coinciding
exactly with the exit of another, while preserving a constant total
number of particles at each discrete simulation time $t_k$.

\subsection{Time Evolution Operator}

In this work, all calculations are performed within the no-quenching
approximation \citep{Kolb58}, which assumes that collisions do not
induce transitions between states with different principal quantum
numbers $n$. As a consequence, the matrix element of the dipole
operator can be expanded over two disjoint sets of upper and lower
states, $|a'\ket$ and $|b'\ket$, which include all states sharing the
same principal quantum number (and spin) as the states $|a\ket$ and
$|b\ket$ involved in the radiative transition under consideration:
\begin{equation}
\bra b|{\bf d}(t)|a\ket = \sum_{a'b'} \bra b|U_b^\dag(t,0)|b'\ket
\bra b'|{\bf d}|a'\ket \bra a'|U_a(t,0)|a\ket.
\label{eqn:bda_HeI}
\end{equation}

\noindent
Here, ${\bf d}$ denotes the dipole moment operator in the Schrödinger
picture, while $U_a$ and $U_b$ are the time evolution operators
projected onto the upper and lower subspaces of states, respectively.

The projected evolution operators satisfy the equations
\begin{eqnarray}
i\hbar {dU_a(t,0)\over dt} &=& H_a(t)\  U_a(t) \nonumber \\
i\hbar {dU_b(t,0)\over dt} &=& H_b(t) \ U_b(t)
\label{eqn:UaUb_HeI}
\end{eqnarray}  

\noindent
where $H_a$ and $H_b$ are the projections of the total Hamiltonian
(Equation~\ref{eqn:H_HeI}) onto the corresponding subspaces. This
formalism explicitly incorporates lower-state mixing in the
simulations. While this effect was already included in earlier studies
\citep{Gigosos96,Gigosos2014b,Cho2022}, it was neglected in the calculations of
\citet{PTremblay2020}.

Once a temporal sequence of electric fields has been generated at
$N_t$ consecutive time steps $t_k \equiv k\Delta_t$ (with $k$ ranging
from 0 to $N_t-1$), the time evolution operators $U_a(t_k,0)$ and
$U_b(t_k,0)$ of the perturbed atom must be evaluated in the
corresponding upper and lower subspaces. Following \citet{Gigosos96},
we assume that the electric field—and therefore the Hamiltonian
operator $H$—remains constant within each time interval
$(t_{k-1},\,t_{k})$. Under this approximation, the operators $U_a$ and
$U_b$ are diagonal in the basis of their respective instantaneous
eigenstates:
\medskip
\begin{eqnarray}
\langle \alpha' |U_a(t_k, t_{k-1})|\alpha\rangle
&=&
e^{-iE_\alpha \Delta_t/\hbar} \delta_{\alpha\alpha'}
\nonumber \\
\langle \beta' |U_b(t_k, t_{k-1})|\beta\rangle
&=&
e^{-iE_\beta \Delta_t/\hbar} \delta_{\beta\beta'}.
\end{eqnarray}

\noindent
Here, $E_{\alpha}$ and $E_{\beta}$ denote the energies of the
perturbed upper and lower states $|\alpha\ket$ and $|\beta\ket$,
respectively. This property allows the matrix elements of the
operators $U_a(t_k,\,t_{k-1})$ and $U_b(t_k,\,t_{k-1})$ to be
expressed in the basis of unperturbed states. For the upper states,
one obtains
\begin{equation}
\label{eq:SchrodU6}
\begin{aligned}
\langle a'|U_a(t_k, t_{k-1})|a\rangle
&= \sum_{\alpha}
e^{-iE_{\alpha}\Delta_t/\hbar}\langle a'|\alpha\rangle \langle \alpha | a\rangle ,
\end{aligned}
\end{equation}

\noindent
where the perturbed-state energies $E_\alpha$ and the transformation
coefficients $\bra \alpha | a \ket$ are obtained by solving the
corresponding eigenvalue problem; an analogous expression applies to
the lower states.

The full evolution operators $U_a(t_k,\,0)$ and $U_b(t_k,\,0)$ are
then constructed iteratively by matrix multiplication:
\begin{equation}
\label{eq:SchrodU630}
\begin{aligned}
\bra a'|U_a(t_k, 0)|a\ket
&=\\
&\sum_{a''}\bra a'| U_a(t_k, t_{k-1})|a''\ket\
\bra a''| U_a(t_{k-1}, 0)|a\ket \ \\
\bra b|U_b(t_k, 0)|b'\ket
&=\\
&\sum_{b''}
\bra b| U_b(t_k, t_{k-1})|b''\ket\
\bra b''| U_b(t_{k-1}, 0)|b'\ket.
\end{aligned}
\end{equation}

\noindent
This numerical integration scheme preserves the unitarity of the time
evolution operator by construction.

The validity of this procedure relies on the assumption that the
electric field varies negligibly over a single time step
$\Delta_t$. The time step is therefore chosen as a fraction $\epsilon$
of the characteristic timescale associated with the temporal variation
of the electric field,
\begin{equation}
\label{eq:deltt}
\Delta_t= \epsilon\frac{r_0}{v_T},
\end{equation}

\noindent
where $r_0$ denotes the typical interparticle distance in a plasma of
electron density $N_e$, and $v_T$ is the electron thermal velocity,
computed using the reduced mass $\mu_e$. Typical values of $\epsilon$
used in previous studies are $\epsilon=0.01$
\citep{Gigosos96,PTremblay2020} and $\epsilon=0.05$ \citep{Cho2022}.

\subsection{Implementation Details \label{subsec:ID}}

The simulation method is computationally demanding, which makes the
careful selection of its parameters essential. Choosing these
parameters requires a delicate balance between achieving accurate line
profiles and maintaining reasonable computational time. The main
parameters of the model include the time step, controlled by the
parameter $\epsilon$ (see Equation~\ref{eq:deltt}), the total number
of time steps $N_t$, the radius of the simulation volume $R$,
expressed as a multiple $K_D$ of the Debye radius, 
\begin{equation}
R = K_D \sqrt{\frac{k T}{4 \pi  (N_e + N_{\rm He\,II})\, e^2}},
\end{equation}
and the number of independent time sequences of electric fields used
to compute the final profile.

In previous work \citep{PTremblay2020}, the simulation volume and the
calculation volume were treated as separate entities. The simulation
volume contained all moving perturbers, while only those particles
located within a smaller calculation volume contributed to the
electric field at the position of the emitting atom.  However, this
approach introduces a significant computational overhead, especially
at low electron densities. In the present study, we simplify the
procedure by equating the simulation and calculation volumes. This
modification eliminates unnecessary computations without compromising
accuracy.

The choice of the multiple $K_D$ is guided by two considerations: (1)
the microfield distribution obtained from the simulation should
closely match the theoretical microfield distribution of
\citet{Hooper68}, and (2) the temporal evolution of the electric field
at the emitter must remain continuous when a perturber leaves the
simulation volume and is replaced. Through extensive testing, we
determined that $K_D = 3$ provides satisfactory results for electron
densities below $10^{17}$ cm$^{-3}$, while $K_D = 5$ is more
appropriate for higher densities. Using a single value for all
densities is problematic because, paradoxically, the number of
perturbers within a sphere of fixed radius (in units of the Debye
length) increases as the density decreases, leading to unnecessarily
long computational times at low densities. Choosing an appropriate
$K_D$ ensures both physical accuracy and computational efficiency
across the full range of densities considered.

Another key parameter is the exclusion radius, which defines the
minimum allowed impact parameter in order to prevent unphysically
large electric fields at the emitter. While one could, in principle,
define the exclusion radius individually for each transition as the
orbital radius of its upper energy levels, we adopt a uniform
exclusion radius for all simulations. Specifically, we choose a value
corresponding to the orbital size of the $n=2$ level of the helium
atom. This choice provides a reasonable balance between accuracy and
computational cost, while preventing extreme field contributions that
could destabilize the simulation.

The time step $\Delta_t$ and the total number of steps $N_t$ must also
be carefully selected. The initial goal is to determine values that
are suitable for all spectral lines and thermodynamic conditions under
consideration. Analysis of the autocorrelation function of the dipole
moment, $C(t)$, indicates that the product $N_t \epsilon$ must be
chosen based on the narrowest spectral lines and the lowest electron
densities in the grid.  Based on tabulated Lorentzian widths computed
using the standard theory (for example, \citealt{Griem74}),
the He~{\sc i} $\lambda4713$ line is significantly narrower than other
lines studied. Therefore, we divide the lines into two categories:
He~{\sc i} $\lambda4713$ and broader lines.

For the broader lines, the value $N_t \epsilon = 2000$ was adopted,
determined from the autocorrelation function of the He~{\sc i}
$\lambda4471$ line at the minimum electron density of $10^{14}\ {\rm
  cm}^{-3}$. This value ensures that the autocorrelation function has
essentially decayed to zero at the end of the simulation duration, $T
= N_t \Delta_t(\epsilon)$, guaranteeing convergence of the line
profiles when computed using the power spectrum method.

The separate choice of $N_t$ and $\epsilon$ is then made with
computational efficiency in mind. Following the approach of
\citet{PTremblay2020}, setting $\epsilon = 0.01$ would lead to $N_t =
200{,}000$, which is computationally expensive. We therefore adopt
$\epsilon = 0.02$, which reduces $N_t$ to 100{,}000 while still
maintaining adequate temporal resolution of the electric field
variations. This choice represents a reasonable compromise between
computational efficiency and the accuracy required to resolve the
narrowest features of the spectral lines.

\begin{figure}
\centering
\includegraphics[width=\columnwidth]{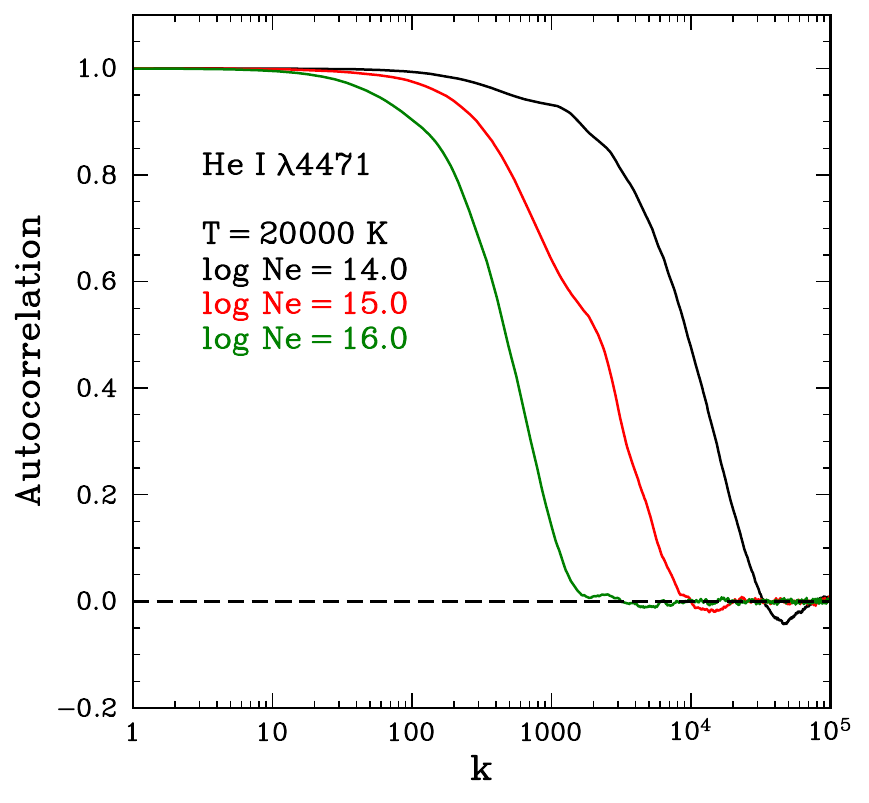}
\caption{Autocorrelation function of He {\sc i} $\lambda$4471 for $T =
  20,000$~K and electron densities $10^{14}$, $10^{15}$, and
  $10^{16}$~cm$^{-3}$ over the simulation time-step index $k=t_k/\Delta_t$.}
\label{fig:Ct4471}
\end{figure}

Figure~\ref{fig:Ct4471} illustrates the autocorrelation function of
the He~{\sc i} $\lambda4471$ line at $T = 20,000$~K for three
different electron densities, plotted as a function of the logarithm
of the time-step index $k$ (defined as $t_k \equiv k\Delta_t$ with $k$
ranging from 0 to $N_t-1$). As implied by Equation \ref{eq:stark9},
computing the line profile by integrating the autocorrelation function
up to the finite duration of the simulation $T$ is only valid if the
autocorrelation function has effectively converged to zero over the
integration domain. The characteristic width of the autocorrelation
function increases as the electron density decreases, reflecting the
slower decay of correlations at lower densities. This behavior imposes
a lower limit on the density for which a given number of time steps is
sufficient.

\begin{figure}
\centering
\includegraphics[width=\columnwidth]{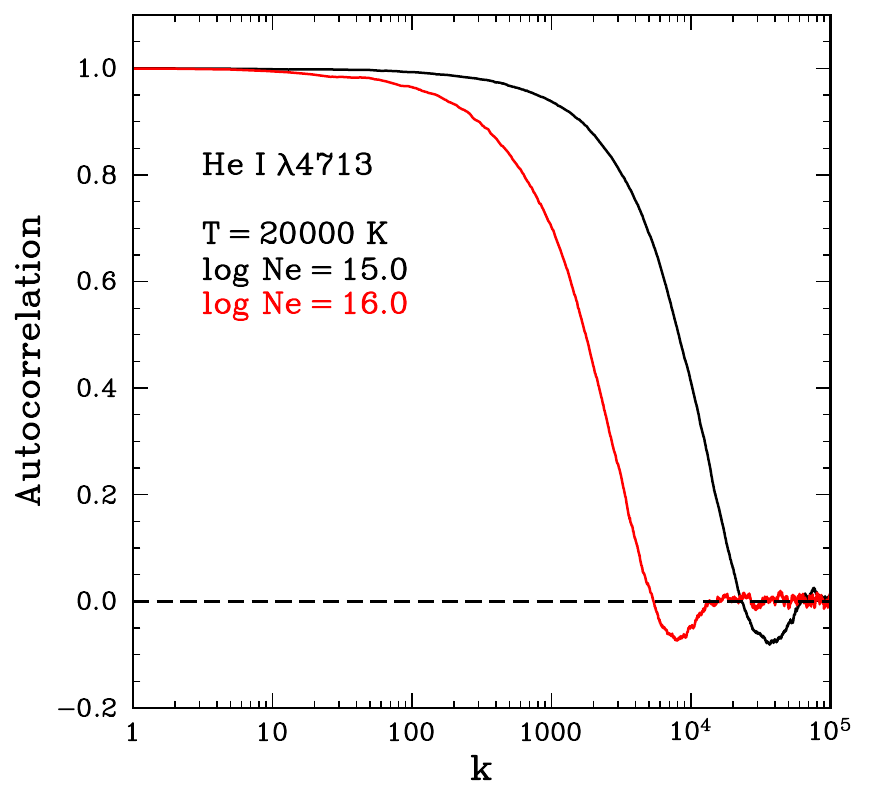}
\caption{Same as Figure \ref{fig:Ct4471} but for 
He {\sc i} $\lambda$4713 for $T =
  20,000$~K and electron densities $10^{15}$ and $10^{16}$~cm$^{-3}$.}
\label{fig:Ct4713}
\end{figure}

For the narrowest line, He~{\sc i} $\lambda4713$, simulations at $N_e
= 10^{14}\ {\rm cm}^{-3}$ would require prohibitive simulation
durations. Consequently, we only compute its profiles for $N_e \ge
10^{15.5}\ {\rm cm}^{-3}$, using the same $N_t$ and $\epsilon$ values
as for broader lines. Figure~\ref{fig:Ct4713} shows the
autocorrelation function for this line at two electron densities.
Three additional narrow lines, He {\sc i} $\lambda\lambda 3889$, 5016,
and 5876 were initially considered, but were excluded due to
computational constraints.

The computation of the line profile $I_i(\omega)$ for a given
simulation volume proceeds in three main steps. First, the time
sequence of electric fields, ${\bf F}_i(t_k)$, at the center of the
simulation volume is generated from the spatial and temporal
distribution of perturbers. Second, the Schrödinger equation
(Equation~\ref{eqn:UaUb_HeI}) is integrated numerically within the
upper and lower state subspaces, and the evolution operator matrix
elements are computed as in Equation~\ref{eq:SchrodU6}. Finally, the
time-dependent dipole matrix elements (Equation~\ref{eqn:bda_HeI}) are
substituted into Equation~\ref{eqn:Liomega2}, and the integral over
time is performed for each required frequency to obtain the final line
profile.

This implementation strategy ensures that all relevant physical
effects—including ion dynamics, electron-impact broadening, and state
mixing—are correctly captured while maintaining convergence and
numerical stability across the full range of densities and line widths
considered.

In summary, our improved profiles differ from those of
\citet{PTremblay2020} in terms of the use of the power spectrum
method, the particle reinjection procedure, the lower-state mixing,
and the identity between the simulation and calculation volumes.

\section{Results\label{sec:comp0}}

With the simulation environment and computational procedure now fully
established, we turn to an analysis of the resulting line profiles.

All He {\sc i} spectral lines considered in this work arise from
transitions between upper levels $|n\ell_a m_as\ket$ with $3 \le n \le
6$ and lower levels $|2\ell_b m_bs\ket$ of identical spin $s$, for
given values of $\ell_a$ and $\ell_b$, within the wavelength range
3800--5100~\AA. The transitions examined include $2P^1-nS^1$
($\lambda\lambda5048$, 4438, and 4169), $2S^1-nP^1$ ($\lambda
3965$), $2P^1-nD^1$ ($\lambda\lambda4922$, 4388, and 4144),
$2P^3-nS^3$ ($\lambda\lambda 4713$, 4121, and 3868), and $2P^3-nD^3$
($\lambda\lambda4471$, 4026, and 3820). Most transitions of the form
$2S-nP$ lie in the near-ultraviolet and were therefore excluded from
the present analysis. 

The spectral line profiles are computed at three temperatures:
10,000~K, 20,000~K, and 40,000~K. As discussed in
Section~\ref{subsec:ID}, the narrowest transition, He {\sc i}
$\lambda$4713, constitutes a particular case. For this line, profiles
are calculated over an electron-density range of $\log N_e = 15.5$ to
$17.5$ in steps of 0.5, with an additional calculation performed at $6
\times 10^{17}$ cm$^{-3}$, which matches the highest density tabulated
by \citet{Shamey69} and \citet{Beauchamp97}. All remaining lines are
also computed at lower densities, specifically $\log N_e = 14.0$,
$14.5$, and $15.0$.  The perturber population consists of electrons
and He~{\sc ii} ions in equal proportions. Contributions from He~{\sc
  iii} ions are neglected, as their number density is expected to be
negligible under the thermodynamic conditions characteristic of DB
white dwarf atmospheres and other stars studied in this work.

The comparison between the present set of line profiles and those
available in the literature is carried out in two stages. We first
concentrate on the two transitions for which profiles including ion
dynamics are available, namely He {\sc i} $\lambda\lambda 4471$ and
4922. A more extensive comparison is then performed for all 13 lines
using a new generation of profiles computed following the methodology
of B97, which is based on the standard Stark broadening theory.  For
consistency, the reference profiles have been modified to establish a
common baseline with the present calculations. In particular, the
revised profiles exclude the effects of line dissolution and adopt a
finer wavelength resolution, matching that employed in the numerical
simulations. Although line dissolution is a well-established physical
effect \citep{Seaton90, Tremblay2009, Gomez2017, Cho2022}, it is not
included in the current simulation framework.

Under these conditions, any remaining discrepancies between the
simulated profiles and those derived from standard theory are expected
to originate primarily from the explicit treatment of ion dynamics,
and possibly from differences in the modeling of electron broadening
in the line wings. In the work of B97, an interpolation between the
one-electron approximation and the impact theory is used to describe
electron broadening. To assess the influence of this assumption, two
additional grids of line profiles have been computed within the
present study using the standard theory, alongside those obtained from
the simulations. One grid assumes that the impact approximation
applies over the entire profile, while the other employs an
interpolation between the impact and one-electron regimes.  By
construction, the simulation method is expected to provide a more
natural and self-consistent description of the transition between
these electronic broadening regimes.

Finally, a new wavelength grid has been constructed for each spectral
line to ensure improved coverage of the profile compared to that used
in B97. Particular attention is paid to regions where subtle features
induced by ion dynamics or forbidden components are expected,
especially at low densities where these effects are confined to narrow
wavelength intervals. The same wavelength grid is adopted for a given
transition at all densities and temperatures, thereby facilitating
interpolation in the computation of synthetic spectra. Each grid
consists of carefully selected wavelength points, designed to remain
compatible with prior assumptions over a wide range of physical
conditions, including typical line widths, density-dependent behavior,
and the presence of forbidden components that may give rise to
secondary maxima.

\begin{figure*}
\centering
\includegraphics[width=5.0in]{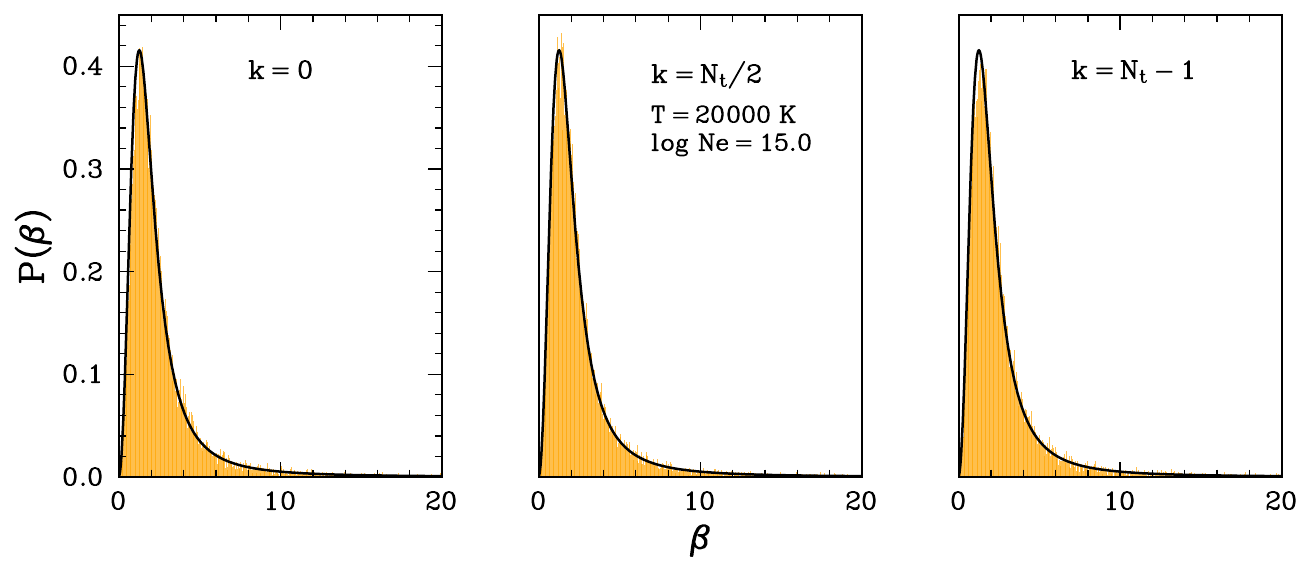}
\caption{Microfield distribution for a temperature of 20,000 K and
  an electronic density of $10^{15}$ cm$^{-3}$ at three time-step
  indices $k=t_k/\Delta_t$ of 0, $N_t/2$ and $N_t-1$. $\beta$ is the intensity of the
  electric field normalized by the Holtzmark field.  The distribution
  from \citet{Hooper68} is shown in black, while the generated
  distribution is represented by the orange histogram.}
\label{fig:hooper10_15}
\end{figure*}

\begin{figure*}
\centering
\includegraphics[width=5.0in]{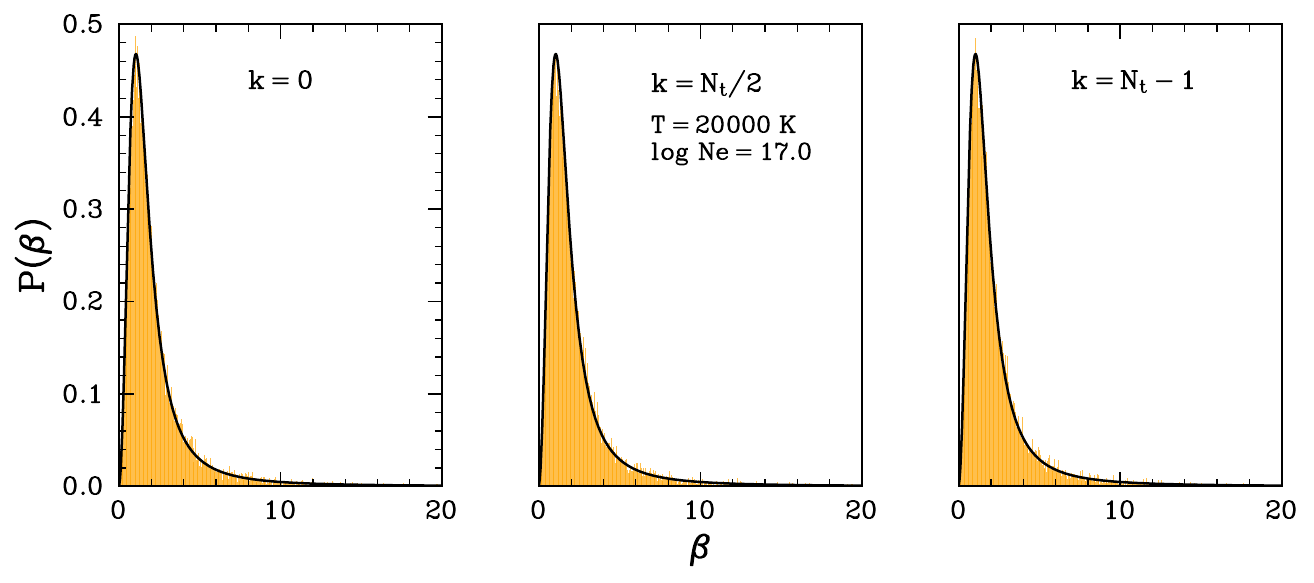}
\caption{Same as Figure \ref{fig:hooper10_15} but for an electronic
  density of $10^{17}$ cm$^{-3}$.}
\label{fig:hooper10_17}
\end{figure*}

\subsection{Statistical Distribution of the Electric Field Intensity }\label{sec:beta}

A standard validation of the simulation environment consists in
comparing the statistical distribution of the electric microfield
generated by the stochastic process with the corresponding theoretical
expectations.

Figures~\ref{fig:hooper10_15} and \ref{fig:hooper10_17} present a
comparison between the ionic electric-field distribution derived by
\citet{Hooper68} and that obtained from the ensemble of simulation
volumes. The comparison is performed at three multiples of the time
step, corresponding to the initial time, the midpoint, and the final
time of the simulation, for two representative electron densities. At
all sampled times, the simulated distributions are found to be
essentially indistinguishable from Hooper’s analytical
distributions. This result demonstrates, first, that the adoption of
the Debye potential successfully reproduces the expected microfield
statistics, in agreement with previous studies \citep{Gigosos96,
  Gigosos2014b, Cho2022}. Second, it confirms that both the
initialization procedure of the simulation volume and the continuous
generation of new particles during the simulation lead to a stationary
stochastic process. The electric-field distributions produced
separately by ions and electrons are likewise found to be stationary.

\begin{figure}[t]
\centering
\includegraphics[width=\columnwidth]{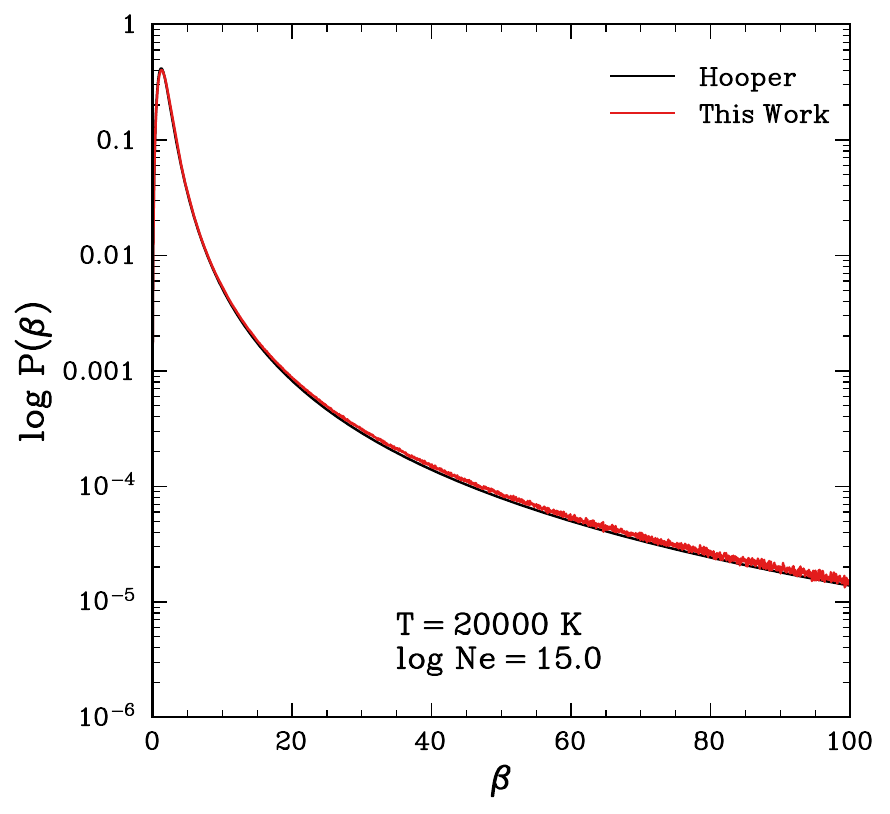}
\caption{Microfield distribution for a temperature of 20,000 K and an
  electronic density of $10^{15}$ cm$^{-3}$ over all time indices.  }
\label{fig:hooper10_15_All}
\end{figure}

Figure~\ref{fig:hooper10_15_All} shows the ionic electric-field
distribution obtained by combining the full set of simulation volumes
over all time indices. This representation provides access to the
behavior of the distribution in the regime of highly improbable
electric-field intensities, corresponding to the asymptotic tail that
contributes to the far wings of spectral lines. As illustrated, the
simulated distribution overlaps almost perfectly with the Hooper
distribution over the entire range, including the low-probability,
high-field domain.

\subsection{Autocorrelation vs. Power Spectrum Methods}\label{sec:CtPS}

\begin{figure}
    \centering
    \includegraphics[width=\columnwidth]{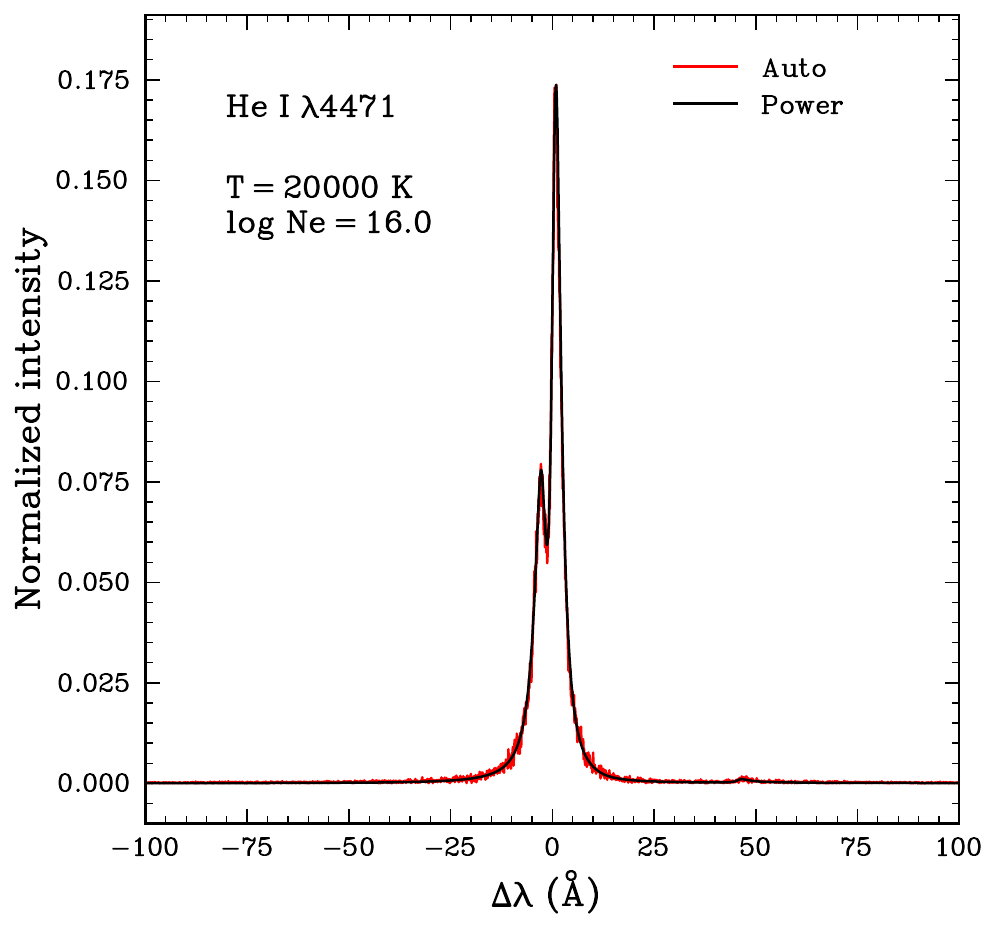}
    \caption{He {\sc i} $\lambda4471$ line profile computed with the
      autocorrelation function and the power spectrum methods, for a
      temperature of 20,000 K and an electronic density of $10^{16}$
      cm$^{-3}$.  }
    \label{fig:prof4471_2methods}
\end{figure}

\begin{figure}
    \centering
    \includegraphics[width=\columnwidth]{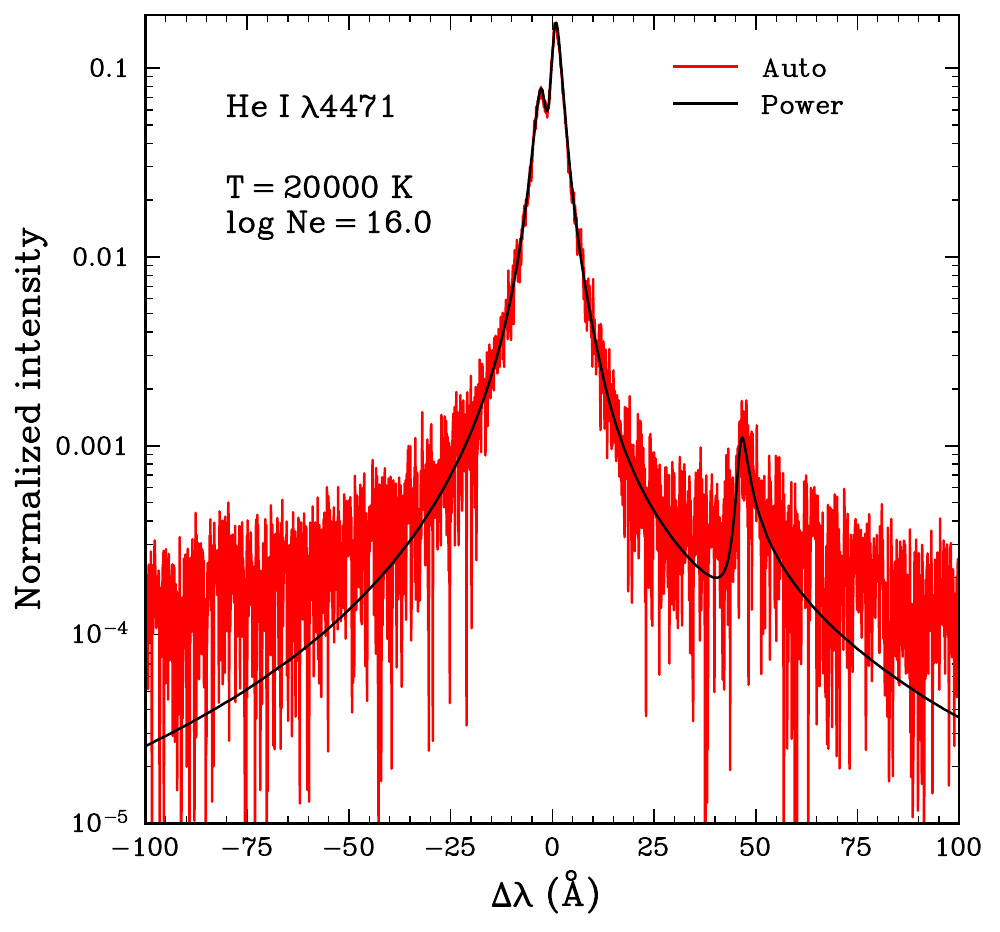}
    \caption{Same as Figure \ref{fig:prof4471_2methods} but on a logarithmic scale.
    }
    \label{fig:prof4471_2methods_log}
\end{figure}

The power spectrum method proves clearly advantageous in terms of
signal-to-noise ratio compared to the method based on the
autocorrelation function, as noted by \cite{Cho2022} for hydrogen
lines. Figure \ref{fig:prof4471_2methods} compares the profile of the
$\lambda4471$ line at a temperature of 20,000 K and an electron
density of $10^{16}$ cm$^{-3}$ obtained using both methods.  While the
signal-to-noise ratio appears satisfactory on a linear scale, a
significant noise-like structure becomes evident when examined on a
logarithmic scale, as shown in Figure
\ref{fig:prof4471_2methods_log}. Such a behavior may introduce
artifacts into the modeled spectra of white dwarfs unless the noise is
removed during a post-processing stage of the line-profile
calculations, given that the strongest spectral lines dominate the
opacity over wavelength intervals extending several tens of Angströms
from their respective line cores.


An analytical model allows for a prediction of an oscillatory
behavior of the line profile computed via the autocorrelation
function.  The model is not intended to quantify the numerical noise
introduced by a finite number of simulation volumes or a discrete time
step. Its more modest objective is to qualitatively assess the effect
of a finite simulation time on the final line profile.  Consider a
spectral line whose profile is Lorentzian in shape, characterized by a
width parameter $a_\omega$ in the angular frequency domain, and
centered at a frequency $\omega_0$.  Its corresponding autocorrelation
function is written as $C(t) = e^{-(a_\omega + i\omega_0)t}$.
Restricting its Fourier transform to a finite time $T$ introduces an
oscillatory term in the profile:
\begin{eqnarray}
I(\omega,T) &=& {1\over \pi} {\rm Re} \int_0^T dt\ e^{-(a_\omega + i\omega_0)t} e^{i\omega t} \nonumber \\
&=& {1\over \pi} {a_\omega \over a_\omega^2 + (\omega - \omega_0)^2} \ (1 + \zeta(\omega))\ ,
\end{eqnarray}

\noindent
which is the product of the desired Lorentzian profile and a term
describing the relative error induced by the finiteness of $T$:

\begin{equation}
\begin{split}
\zeta(\omega) = e^{-a_\omega T}
\Bigg(
&-{\rm cos}\Big((\omega - \omega_0)T\Big)\\
&\quad + {\omega - \omega_0 \over a_\omega}\
{\rm sin}\Big((\omega - \omega_0)T\Big)
\Bigg)\ .
\end{split}
\end{equation}

In the wings of the line, where the deviation from the center greatly
exceeds the characteristic width of the profile ($|\omega - \omega_0|
\gg a_\omega$), the second oscillatory term becomes dominant, allowing
the definition of a $\omega$-dependent amplitude of the oscillations
\begin{equation}
\zeta(\omega) = e^{-a_\omega T}\ {|\omega - \omega_0| \over a_\omega},
\label{eqn:zeta}
\end{equation}
which rises above unity, thereby becoming the primary contributor to the numerical profile.

\begin{figure}[t]
    \centering
    \includegraphics[width=\columnwidth]{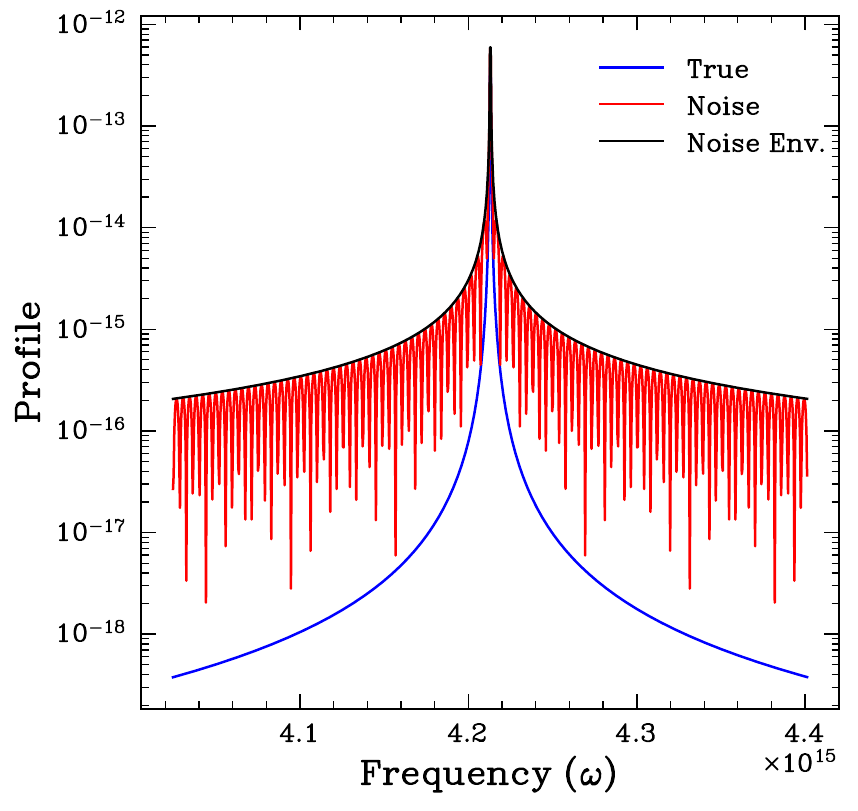}
    \caption{Profile of a generic line with a Lorentzian shape
      computed from truncated Fourier transform of the autocorrelation
      function (see text).}
    \label{fig:profgeneric}
\end{figure}

The Lorentzian shape is not valid in the wings of the $\lambda 4471$
line, where the asymptotic behavior is governed by the linear Stark
broadening due to ions, with $I(\Delta\omega) \propto
(\Delta\omega)^{-5/2}$, in contrast with the Lorentzian form, which
scales as $\Delta \omega^{-2}$.  The linear behavior of the noise
envelope in Equation~(\ref{eqn:zeta}), illustrated in
Figure~\ref{fig:profgeneric} for a generic spectral line, nevertheless
reproduces the amplitude of the oscillations observed in
Figure~\ref{fig:prof4471_2methods_log} for a specific choice of the
model's input parameters.  This does not rule out the possibility that
a similar envelope is present with the power-spectrum method, although
its amplitude appears to be negligible.

\subsection{Normalization of the Profiles from the Standard Theory \label{subsec:norm}}

The comparison between the profile sets obtained from the simulation
and those derived from the semi-analytical method highlighted issues
regarding the normalization of the latter. At low densities, the two
approaches show excellent agreement except at the line cores, where
ion dynamics—absent in the semi-analytical treatment—contribute. The
primary discrepancies, however, arise at high densities, where several
semi-analytical profiles display a systematic downward intensity shift
(on a logarithmic scale) and a consistently smaller integrated
area. This behavior underscores the need for a robust normalization
strategy for semi-analytical profiles.

Two normalization procedures are commonly used in the context of the
semi-analytical theory. The first is numerical normalization, which
consists of integrating the unnormalized numerical profile over a
frequency grid that accurately resolves the underlying continuous
function. The frequency step must be significantly smaller than the
characteristic Stark width near the core of the allowed components and
of the strongest forbidden components, if present, at low density. The
frequency range must also be sufficiently broad to capture the full
profile at high densities. This procedure is applicable to both the
semi-analytical theory and the simulation.

The second approach is specific to the semi-analytical model with
electronic-impact broadening applied over the entire profile, thus
excluding the one-electron description in the wings. In this method,
the numerical profile (Equation \ref{eq:QS2}) is divided by the sum of
squared dipole matrix elements \( \sum_{ab} |\bra a {|\bf d} | b\ket
|^2 \) \citep{Kolb58}. A key advantage of this approach is that the
final frequency grid can be chosen freely without affecting the
normalization.

However, the semi-analytical profiles computed here transition from
electronic-impact broadening near the core to the one-electron theory
in the wings, which invalidates the guarantee of proper normalization
using the above procedure. This hybrid scheme was nonetheless adopted
in the original work (B97), following the methodology of
\citet{Barnard74} and \citet{Barnard75} for the $\lambda\lambda 4471$
and 4922 lines, and of \citet{Gieske69}, who introduced the Lewis
cut-off \citep{Lewis1961} as an alternative but closely related
transition scheme.

In the present study, a second version of the semi-analytical profile
grid was therefore generated using numerical normalization. For each
line, the sampling frequency grid was first validated by computing
profiles broadened solely by electron impacts over the full frequency
range. Across all tested lines, densities, and temperatures, the
resulting profile areas were unity to within 1\%. The final profile
set—used as a benchmark against simulation-derived profiles—combined
electronic-impact and one-electron contributions and was normalized
numerically using the validated grid. It was found that certain lines,
particularly those originating from upper levels with principal
quantum numbers $n = 5$ and 6 at the highest densities, required
numerical normalization. Before normalization, the integrated areas of
these lines were typically about 0.9 and 0.8 at $N_e =
10^{17}$ cm$^{-3}$, respectively. In contrast, the
$\lambda\lambda$4471 and 4922 lines showed negligible differences
between the two normalization methods.

\subsection{Comparison of He {\sc i} $\lambda\lambda$4471 and 4922 \label{sec:4471}}

The validation of the calculated profiles in this work begins with a
detailed comparison against profiles available in the literature. For
neutral helium, numerical simulations providing detailed profiles are
available only for He {\sc i} $\lambda4471$ \citep{Gigosos2009} and
$\lambda$4922 \citep{Lara2012}. These two sets of profile grids cover
electron densities from $10^{15}$ to $10^{18}$ cm$^{-3}$,
approximately corresponding to the upper range of densities in our own
set of profiles.

It is also instructive to compare our profiles with those of
\citet{Barnard74} and \citet{Barnard75}, who account for the
contribution of ion dynamics using a procedure appropriate at low
densities, while providing grids that extend to lower densities, from
$10^{13}$ to $10^{16}$ cm$^{-3}$. These grids therefore
constitute a privileged reference for our profiles at lower densities,
although ion dynamics is not included at their highest density point
($10^{16}$ cm$^{-3}$). The B97 profiles cited in this section are
those specifically regenerated for the present work, using the
normalization procedure described in Section~\ref{subsec:norm}. In all
comparisons, Doppler broadening is applied consistently across
profiles.

\begin{figure*}[p]
\centering
\includegraphics[scale=0.5]{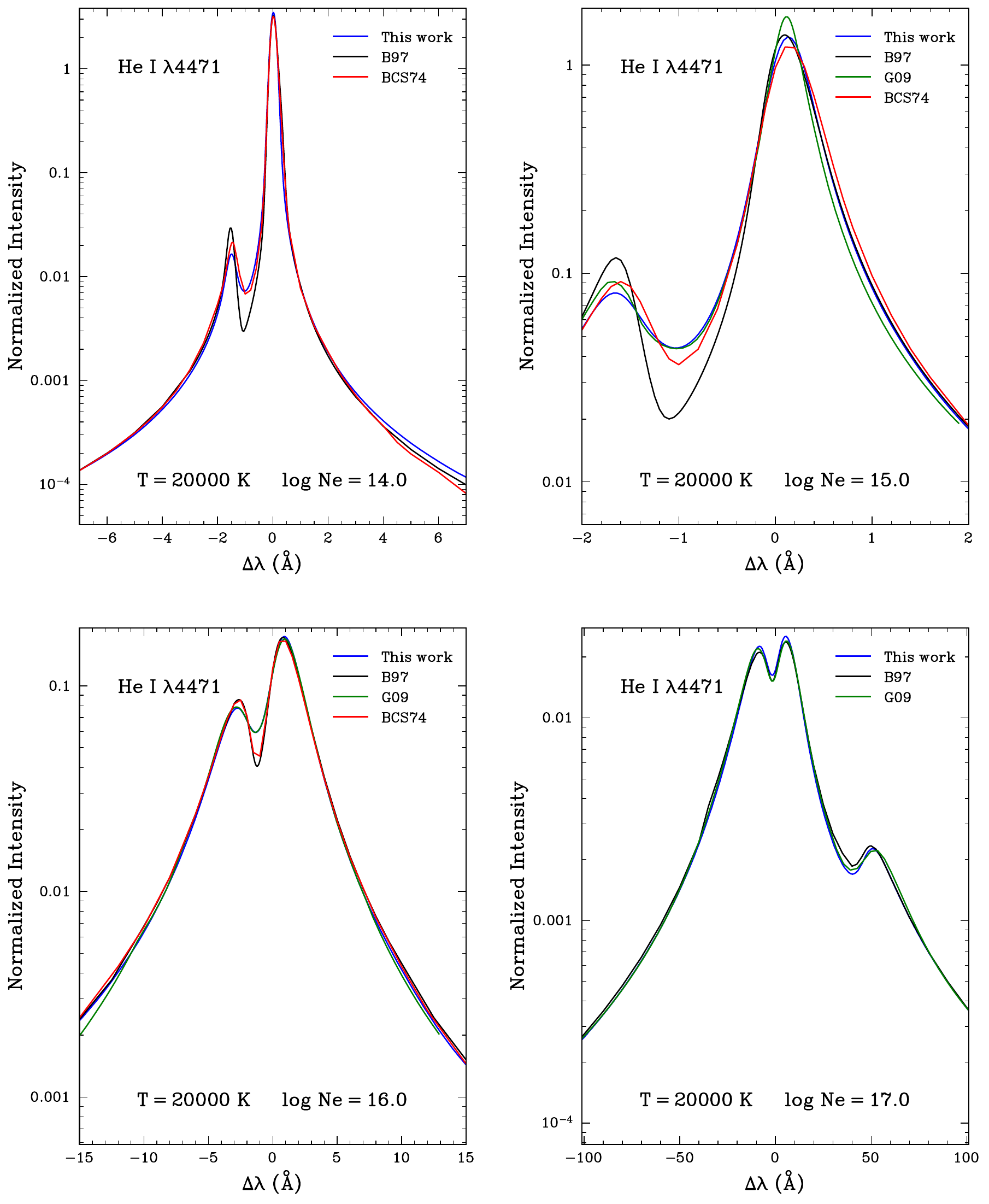}
\caption{He {\sc i} $\lambda$4471 line profile at densities ranging
  from 10$^{14}$ to 10$^{17}$ cm$^{-3}$ and a temperature of 20,000 K,
  from \citet[][BCS74]{Barnard74}, \citet[][G09]{Gigosos2009},
  \citet[][B97]{Beauchamp97}, and this work.}
\label{fig:fit4471}
\end{figure*}

\begin{figure*}[p]
\centering
\includegraphics[scale=0.5]{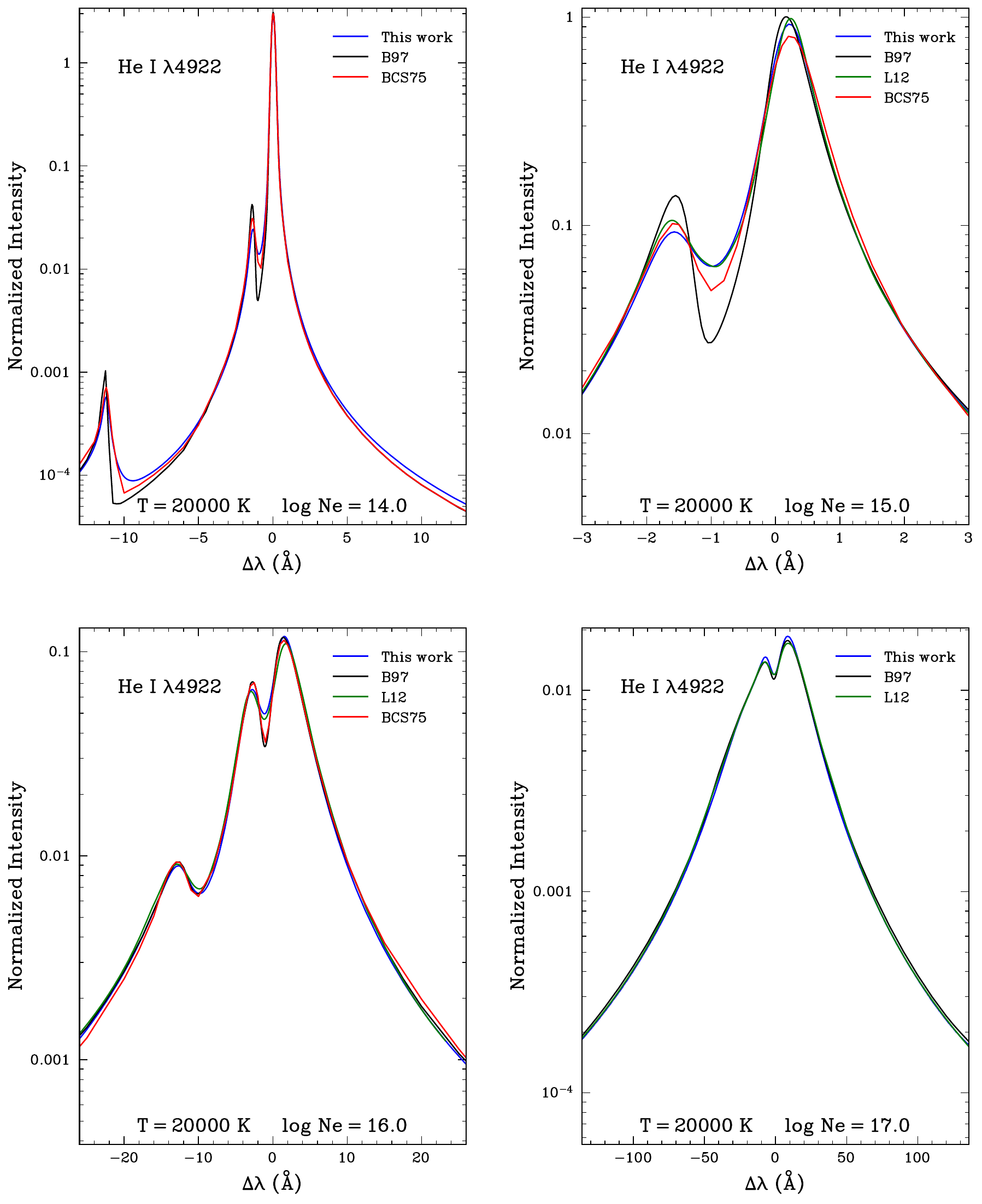}
\caption{He {\sc i} $\lambda$4922 line profile at densities ranging
  from 10$^{14}$ to 10$^{17}$ cm$^{-3}$ and a temperature of 20,000 K,
  from \citet[][BCS75]{Barnard75}, \citet[][L12]{Lara2012},
  \citet[][B97]{Beauchamp97}, and this work.}
\label{fig:fit4922}
\end{figure*}

Figure~\ref{fig:fit4471} compares the $\lambda 4471$ profiles from
these authors with our results at electron densities of $10^{14}$,
$10^{15}$, $10^{16}$, and $10^{17}$ cm$^{-3}$, and a temperature
of 20,000~K. These four densities allow us to capture the
full range of behaviors of interest and to highlight differences among
the profiles reported by various authors. At the lowest density, our
profile shows excellent agreement with \citet{Barnard74}, including
the line core and the region between the allowed and nearest forbidden
components, where ion dynamics contribute—an effect neglected in
B97. The agreement remains excellent at $10^{15}$ cm$^{-3}$, this
time including the results of \citet{Gigosos2009}. At
$10^{16}$ cm$^{-3}$, our simulation profiles and those of
\citet{Gigosos2009} are nearly superimposed, whereas the profile of
\citet{Barnard74} aligns more closely with B97, both of which neglect
ion dynamics. At the highest density, the profiles remain in good
overall agreement, including in the wings, where the transition
between the electronic-impact and one-electron regimes becomes
significant in the standard theory as implemented by B97. Similar
trends are observed for $\lambda 4922$, as shown in
Figure~\ref{fig:fit4922}.

The comparison of $\lambda\lambda 4471$ and 4922 profiles, with and
without ion dynamics, highlights two well-known phenomena at low
densities when ion motion is accurately accounted for. First, the gap
between the allowed component and its nearest isolated forbidden
component is partially filled. Second, the forbidden component
farthest from the line core (when sufficiently excited to be
observable) exhibits a shallower slope on the side facing the allowed
component. These effects are sufficiently pronounced to produce
observable consequences in the spectra of low-gravity stars, where
line formation occurs at densities on the order of $10^{14}$ cm$^{-3}$
(see Sections~\ref{subsec:bar} and~\ref{subsec:hd}).

\begin{figure}[h]
\centering
\includegraphics[width=\linewidth]{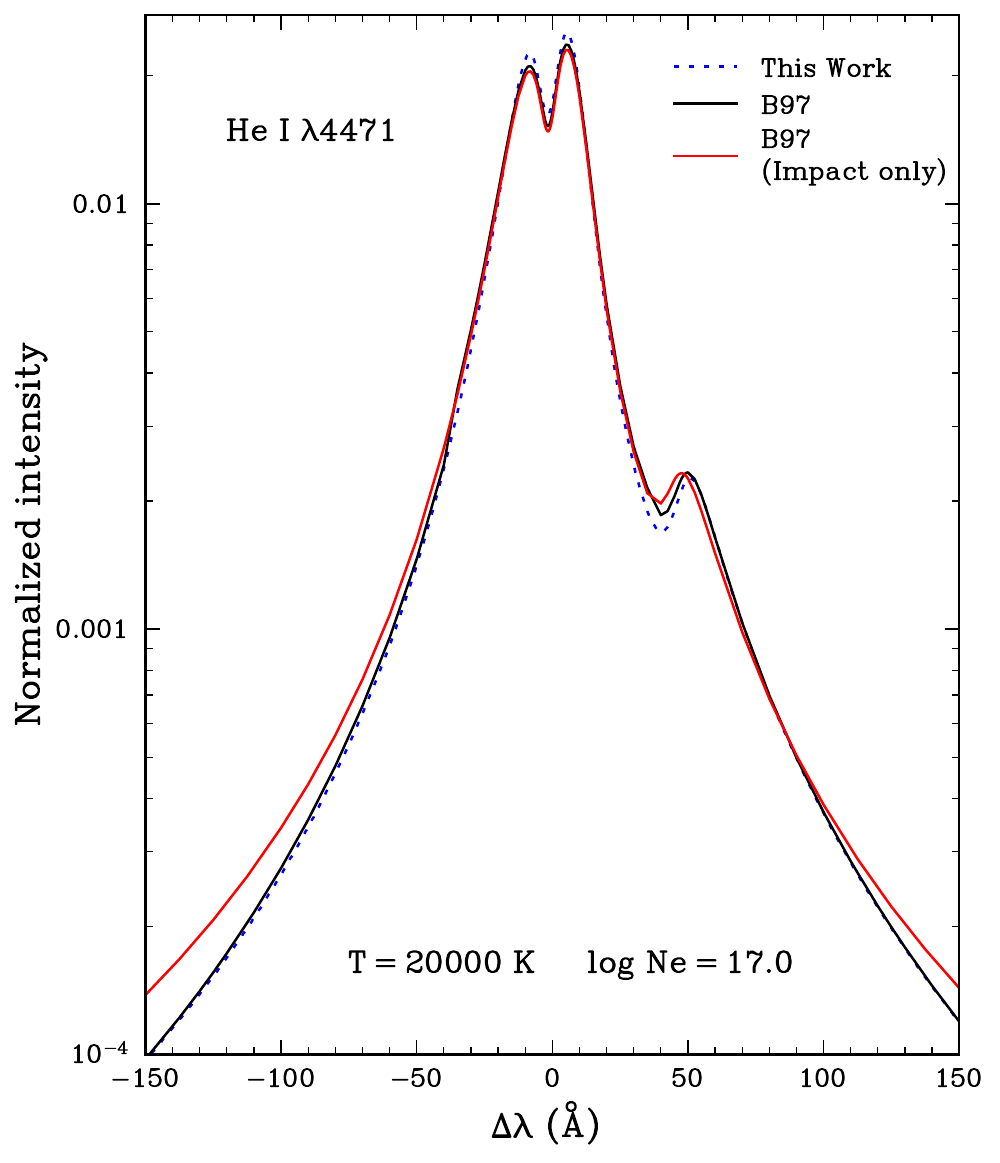}
\caption{He {\sc i} $\lambda$4471 line profile at a density of
  10$^{17}$ cm$^{-3}$ and a temperature of 20,000 K, from
  \citet[][B97]{Beauchamp97} with the semi-analytical theory
    including the impact approximation only as well as both impact and
    one-electron contributions, and this work.}
\label{fig:4471_17_B25}
\end{figure}

To assess the electron contribution in the wings,
Figure~\ref{fig:4471_17_B25} compares He {\sc i} $\lambda 4471$
profiles at an electron density of $10^{17}$ cm$^{-3}$ and 20,000~K
using three approaches: the simulation, the semi-analytical theory
with electronic-impact approximation only, and the semi-analytical
theory including both impact and one-electron contributions. As shown,
the impact theory predicts stronger wings and a shift of the He {\sc
  i} $\lambda4517$ forbidden component, which is not consistent with
the simulation results. This behavior is observed, and often more
distinctly, in the majority of other spectral lines. Although not
rigorously justified, interpolation between the two electron treatment
approaches is therefore preferred, at least for the specific lines and
thermodynamic conditions considered here.

\subsection{Comparison Between Computer Simulations and Standard Theory for All Spectral Lines
\label{subsec:comp_all}}

An overview of the remaining He {\sc i} lines, excluding
$\lambda\lambda4471$ and $4922$, is presented to compare the
predictions of the simulation model with those of the standard theory
using proper normalization and incorporating the transition to the
one-electron approximation in the line wings. All line profiles have
been convolved with a Doppler broadening function.

\begin{figure*}[p]
\centering
\includegraphics[scale=0.5]{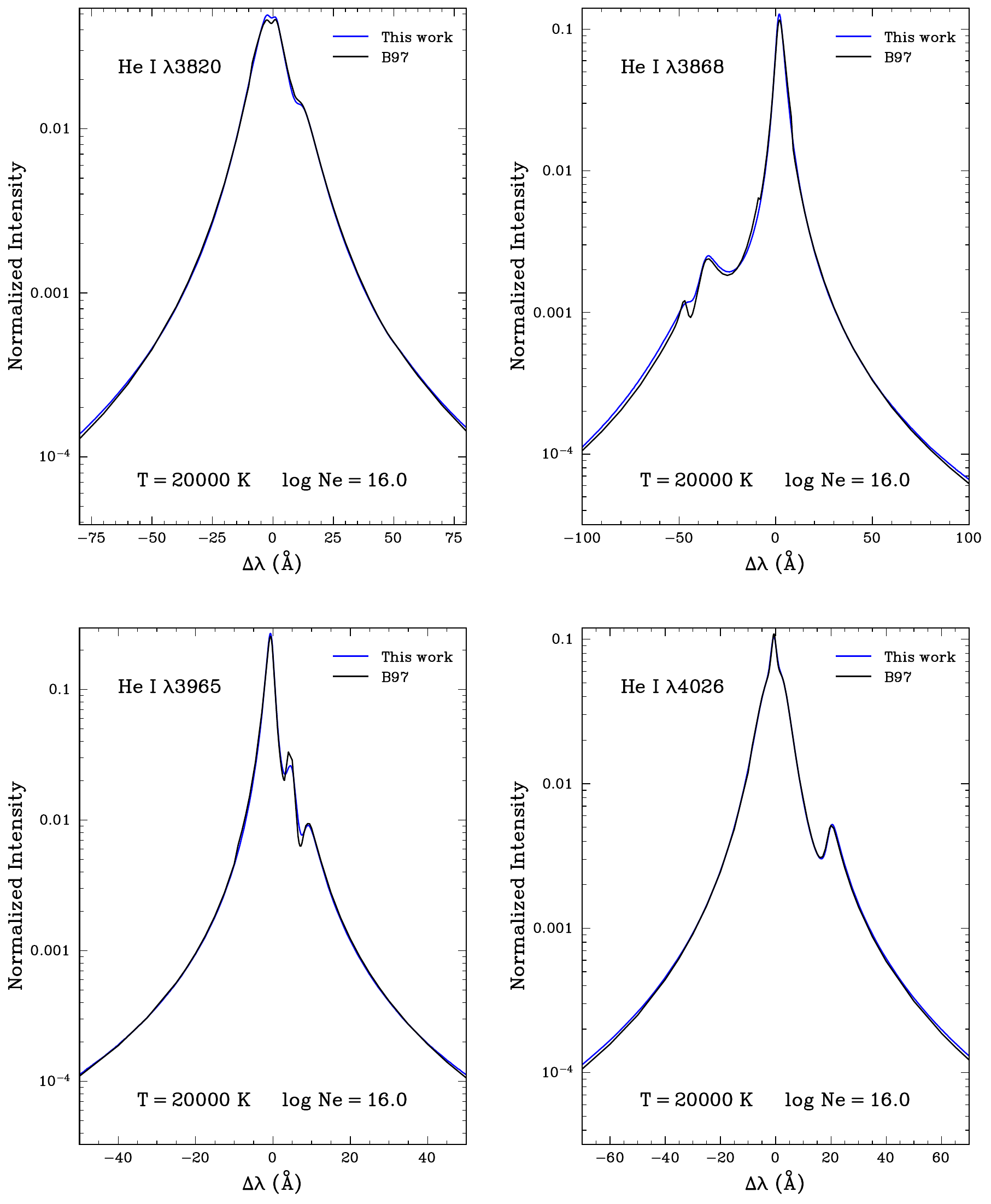}
\caption{He {\sc i} line profiles (indicated in each panel) at an
  electron density of 10$^{16}$ cm$^{-3}$ and a temperature of 20,000
  K, from \citet[][B97]{Beauchamp97} and this work.}
\label{fig:fitAll1}
\end{figure*}

\begin{figure*}[p]
\centering
\includegraphics[scale=0.5]{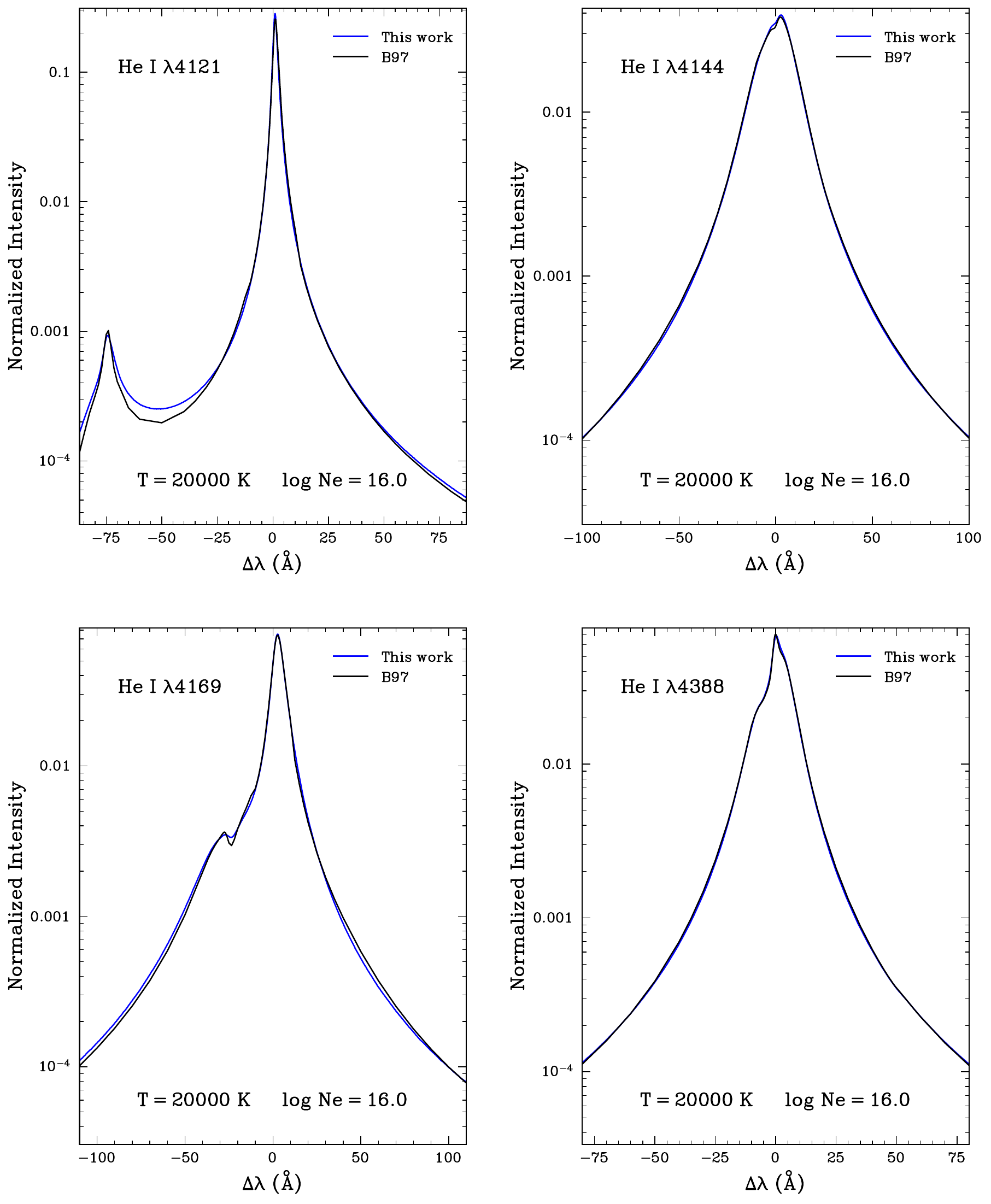}
\caption{Same as Figure \ref{fig:fitAll1}.}
\label{fig:fitAll2}
\end{figure*}

\begin{figure*}[t]
\centering
\includegraphics[scale=0.5]{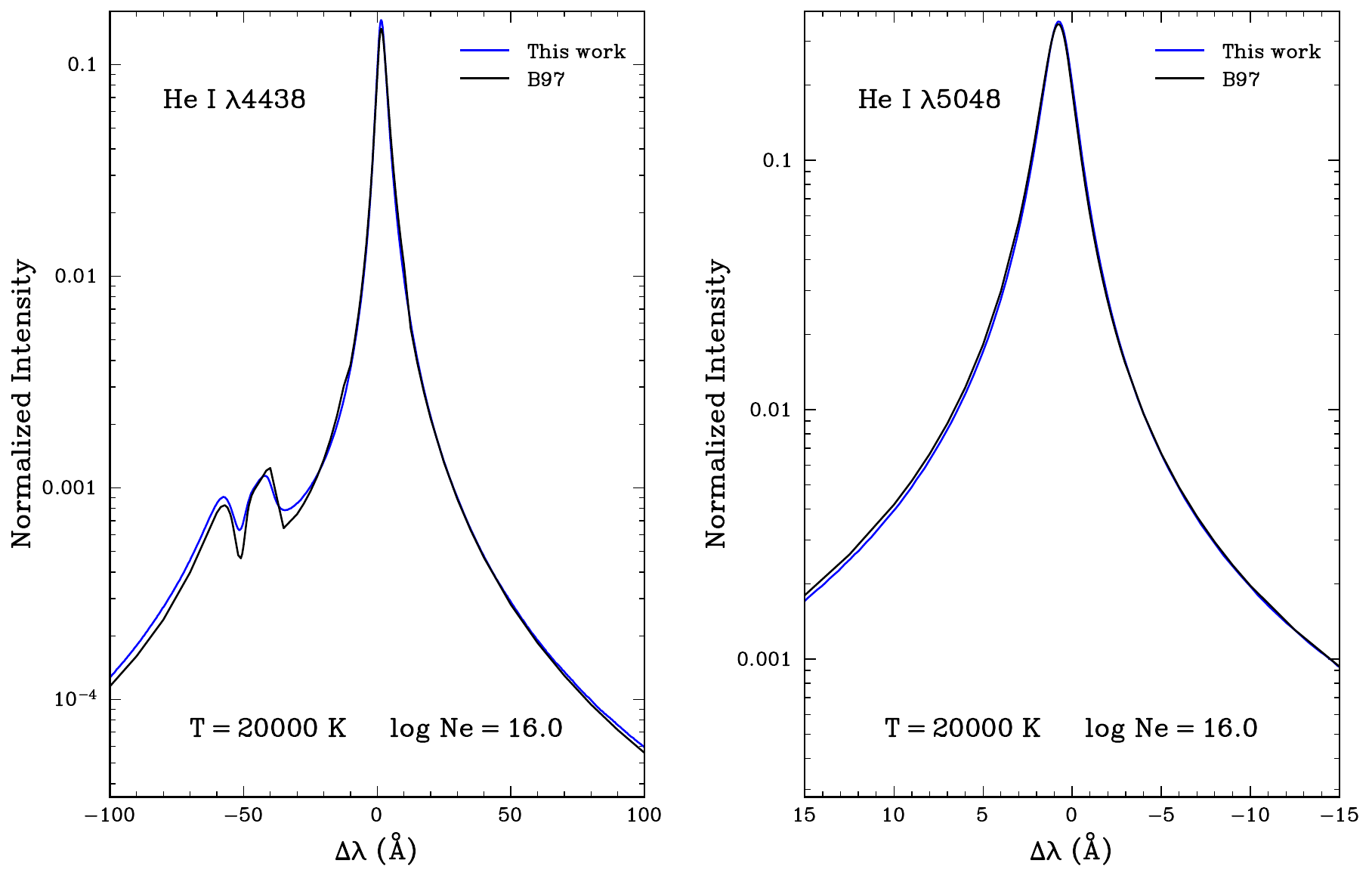}
\caption{Same as Figure \ref{fig:fitAll1}.}
\label{fig:fitAll3}
\end{figure*}

Figures~\ref{fig:fitAll1} to~\ref{fig:fitAll3} display the profiles at
an electron density of $10^{16}$~cm$^{-3}$ and a temperature of
20,000~K. At this temperature and density, the two generations of
profiles are nearly identical.

\begin{figure}[t]
\centering
\includegraphics[width=0.9\linewidth]{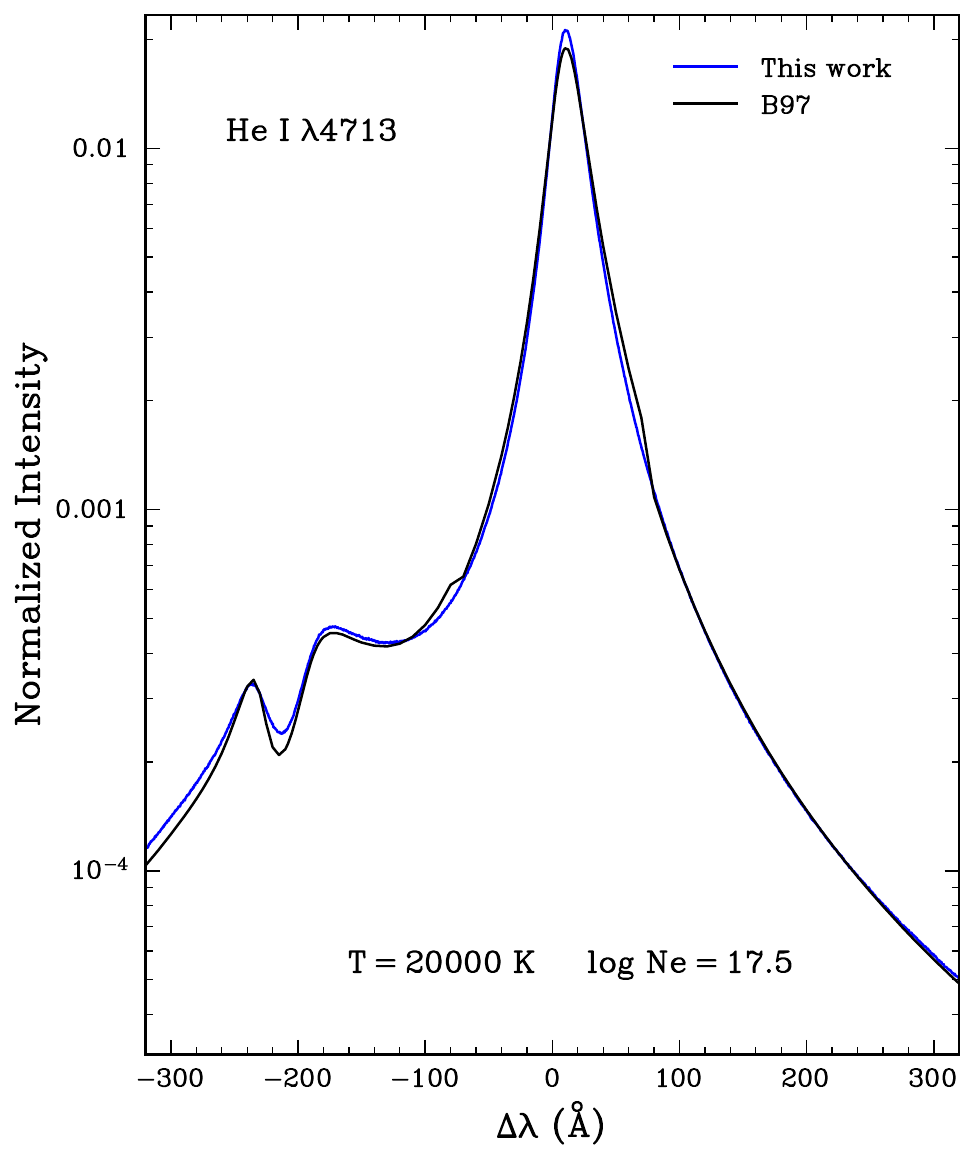}
\caption{He {\sc i} $\lambda$4713 line profile at an electron density
  of $10^{17.5}$ cm$^{-3}$ and a temperature of 20,000 K, from
  \citet[][B97]{Beauchamp97} and this work.}
\label{fig:4713}
\end{figure}

The comparison of high-density profiles allows identification of
limitations in the transition procedure between the electronic-impact
and one-electron regimes within the semi-analytical method as
implemented in B97, and may provide guidance for future
corrections. Figure~\ref{fig:4713} compares the profile of $\lambda
4713$ at a density of $10^{17.5}$~cm$^{-3}$ and 20,000~K. The B97
profile exhibits a discontinuity in the derivative at approximately
$80\ \text{\AA}$ from the line core on both sides. As discussed in
B97, their interpolation scheme between electronic regimes uses
non-differentiable functions (absolute value and Max functions). In
contrast, the profiles obtained from numerical simulations naturally
transition between regimes and could serve as a reference for
developing a more robust method in the future.

An alternative representation of the compared profiles is provided by
a heat map, which illustrates the relative differences between the two
sets of profiles across the full range of densities at constant
temperature. This visualization highlights broader systematic effects
that may be overlooked when examining individual profiles at specific
densities.

\begin{figure*}
\centering
\includegraphics[width=3.2in, clip=true, trim=0.in 0.in 0.in 0.0in]{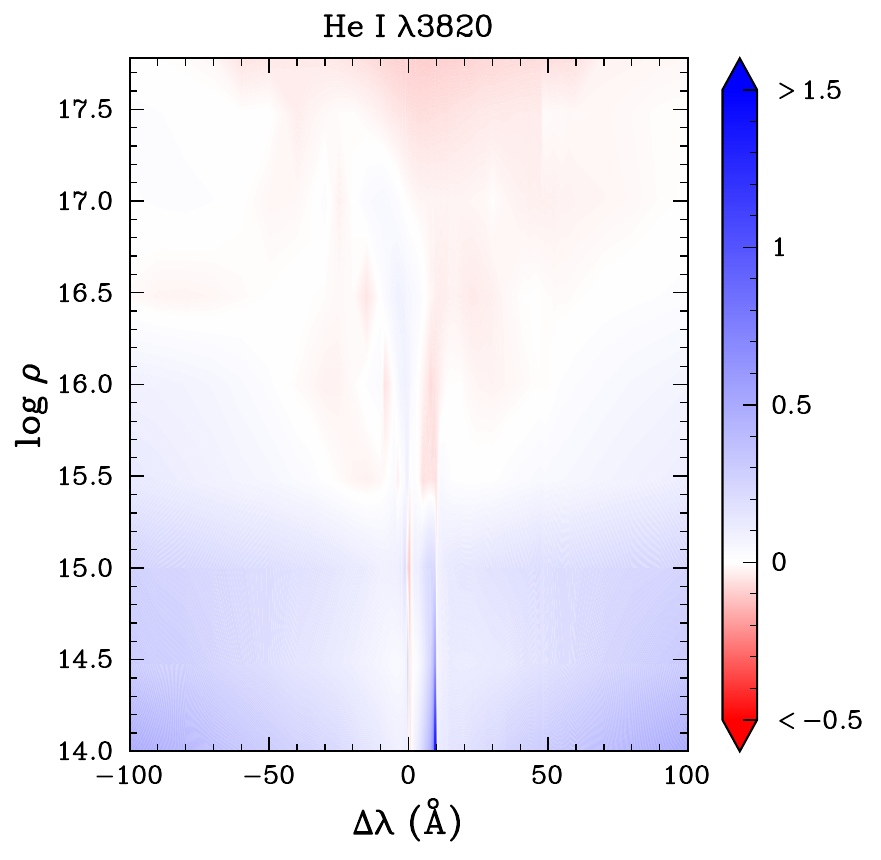}
\includegraphics[width=3.2in, clip=true, trim=0.in 0.in 0.in 0.0in]{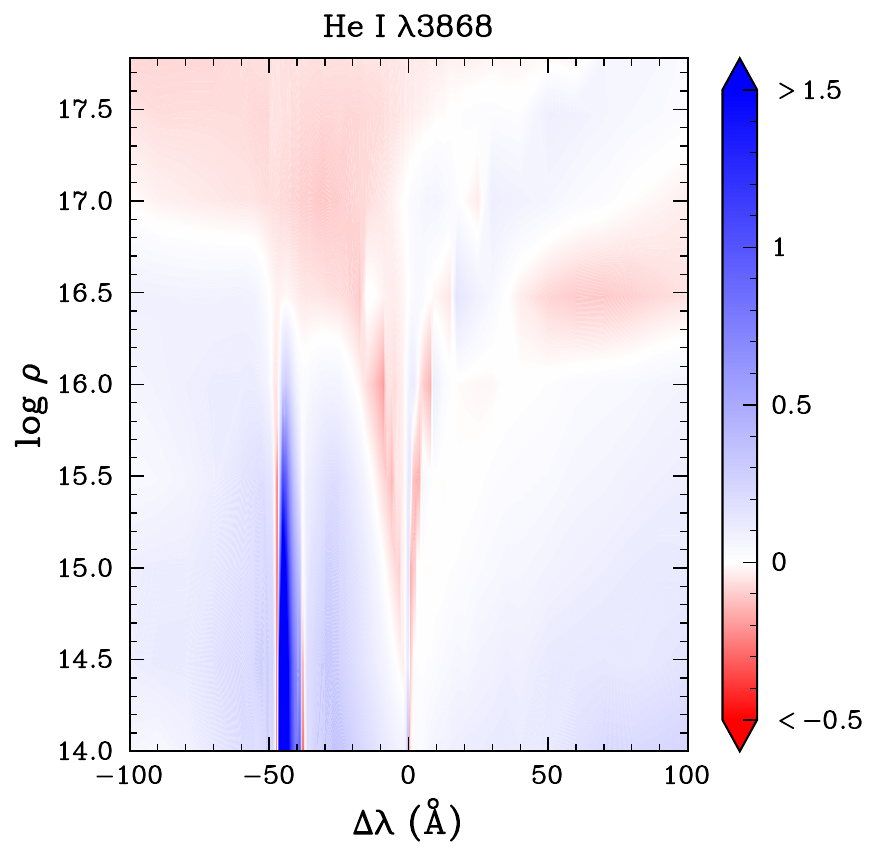}
\includegraphics[width=3.2in, clip=true, trim=0.in 0.in 0.in 0.0in]{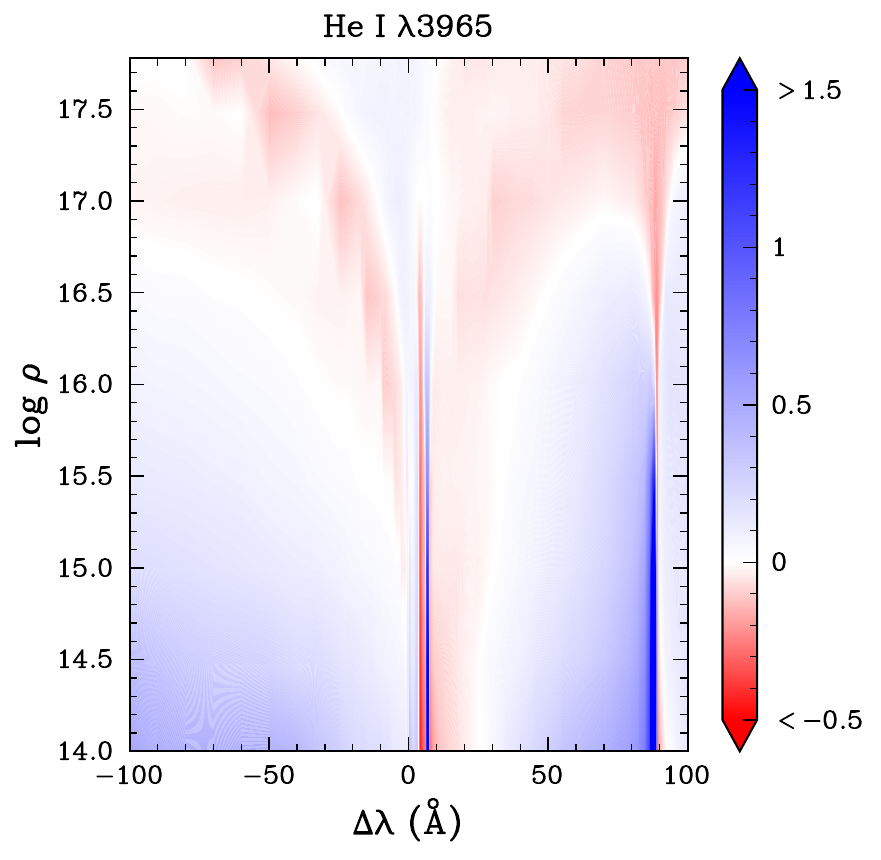}
\includegraphics[width=3.2in, clip=true, trim=0.in 0.in 0.in 0.0in]{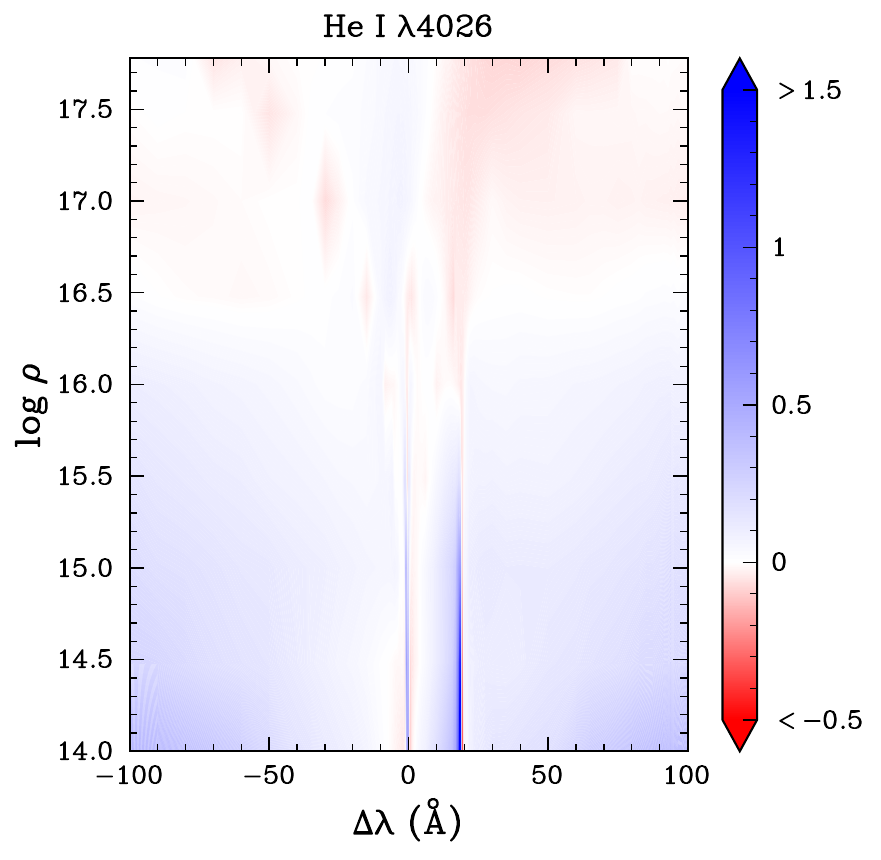}
\caption{Heat map (see text) for He {\sc i} $\lambda\lambda$3820,
  3868, 3965 and 4026 at electron densities ranging from 10$^{14}$ to
  $6\times10^{17}$ cm$^{-3}$ and a temperature of 20,000 K, from
  \citet[][B97]{Beauchamp97} and this work. }
\label{fig:heatmap1}
\end{figure*}

\begin{figure*}
\centering
\includegraphics[width=3.2in, clip=true, trim=0.in 0.in 0.in 0.0in]{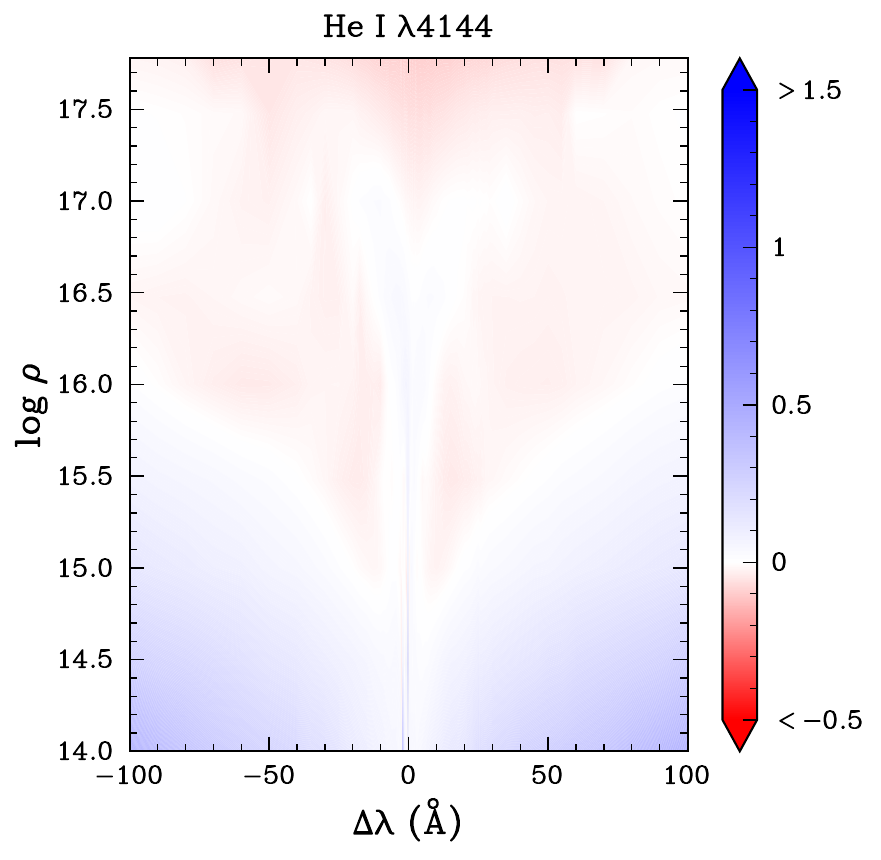}
\includegraphics[width=3.2in, clip=true, trim=0.in 0.in 0.in 0.0in]{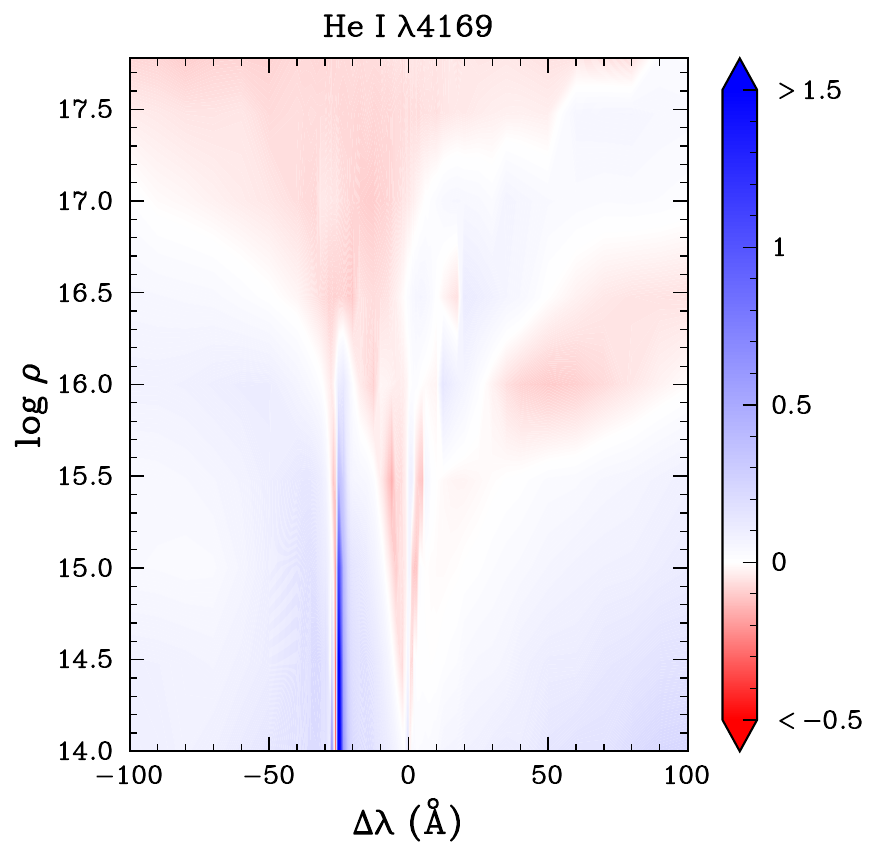}
\includegraphics[width=3.2in, clip=true, trim=0.in 0.in 0.in 0.0in]{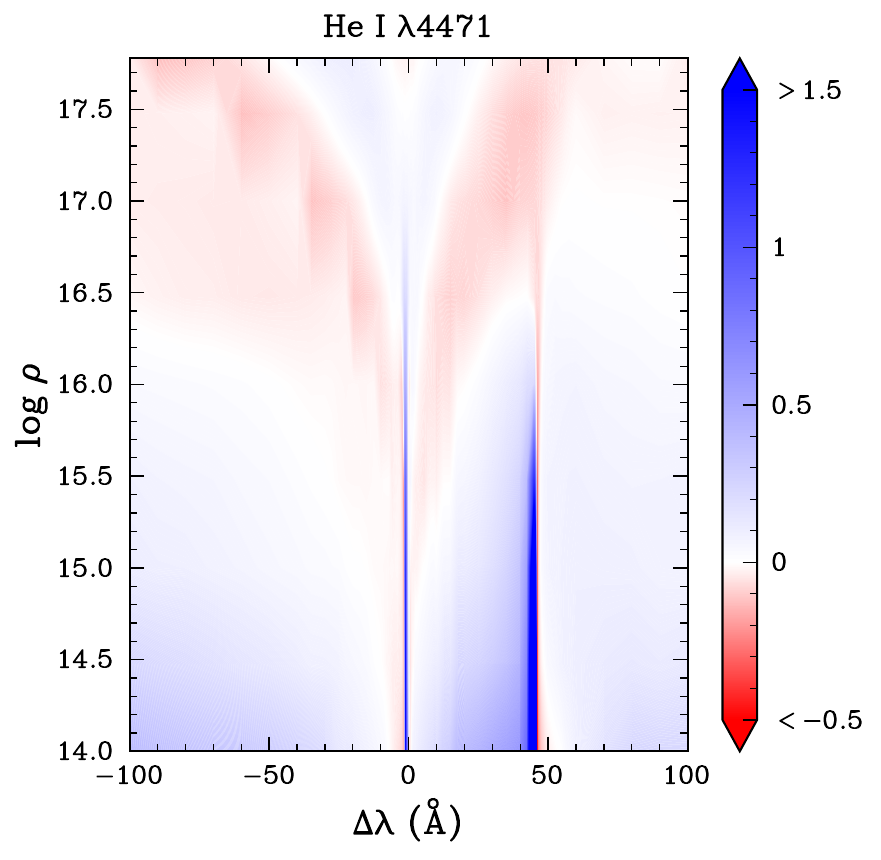}
\includegraphics[width=3.2in, clip=true, trim=0.in 0.in 0.in 0.0in]{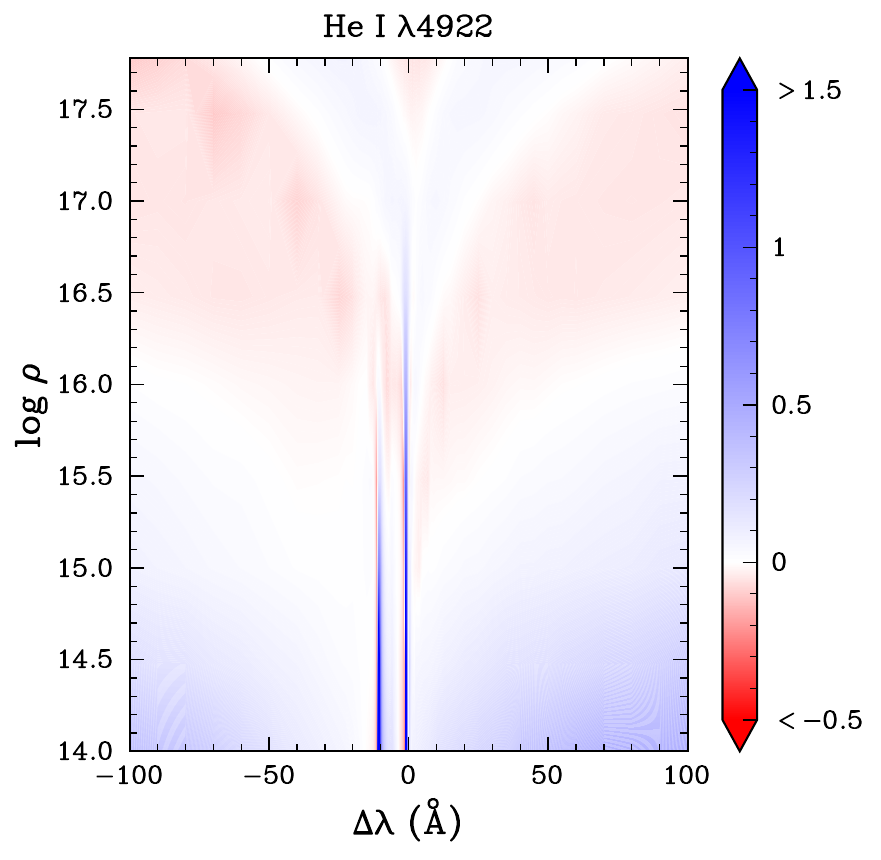}
\caption{Same as Figure \ref{fig:heatmap1} but for He {\sc i} $\lambda\lambda$4144, 4169, 4471 and 4922.}
\label{fig:heatmap2}
\end{figure*}

Figures~\ref{fig:heatmap1} and~\ref{fig:heatmap2} employ a color scale
to encode the relative difference: blue indicates that the line
intensity at a given wavelength is greater in the profile obtained in
this work compared to B97, whereas red indicates the opposite. The
range of red intensities is configured to be narrower than that of the
blue intensities.

As discussed previously, semi-analytical profiles often display
near-discontinuities around distant forbidden
components. Incorporating ion dynamics mitigates these irregularities
by broadening the affected regions. All lines consequently exhibit
narrow blue strips, either near the core or adjacent to forbidden
components, frequently paired with a red band. This pattern reflects a
redistribution effect required to approximately conserve the area of
each component. Focusing on the forbidden components, substantial
differences persist up to densities of $10^{16.5}$ cm$^{-3}$, at which
point the effect of ion dynamics becomes negligible. A similar trend
is observed for the core components of certain lines, notably $\lambda
4471$ and $\lambda 4922$.

\section{Applications}\label{sec:appl}

In this section, we present several applications of our new
Stark-broadening calculations for neutral helium lines. The first
concerns white dwarf stars, for which the semi-analytical line
profiles from B97 have been widely adopted. Because these are
high-gravity objects, the improved treatment of ion dynamics is not
expected to produce a significant change in the line
profiles. However, ion dynamics is expected to play an important role
in the next two low-$\logg$ objects: the ultraviolet-bright star
Bernard 29 in the globular cluster M13 \citep{Dixon2019}, and the
extreme helium-strong star HD 144941 \citep{Przybilla2021}.

\subsection{White Dwarf Spectra}

The semi-analytical He~{\sc i} Stark profiles from B97 have been
employed in several comprehensive spectroscopic analyses of large
samples of DB white dwarfs, notably in the Palomar-Green
\citep{Bergeron2011} and SDSS \citep{Genest2019b} surveys. In these
studies, the atmospheric parameters—effective temperature, surface
gravity, and hydrogen-to-helium abundance ratio—are derived using the
spectroscopic technique, in which observed optical spectra are matched
to model predictions through a $\chi^2$ minimization procedure (see
\citealt{Bergeron2011} for details). The model spectra are obtained by
solving the radiative transfer equation using thermodynamic structures
(temperature and pressure as functions of optical depth) provided by
model atmosphere calculations. As a result, the synthetic spectra
sample a broad range of physical conditions throughout the stellar
atmosphere, in contrast to the illustrative examples presented above,
which correspond to single values of temperature and electron density.

\begin{figure}[h]
\centering
\includegraphics[clip=true,trim=1cm 3cm 1.0cm 4.0cm,width=\columnwidth]{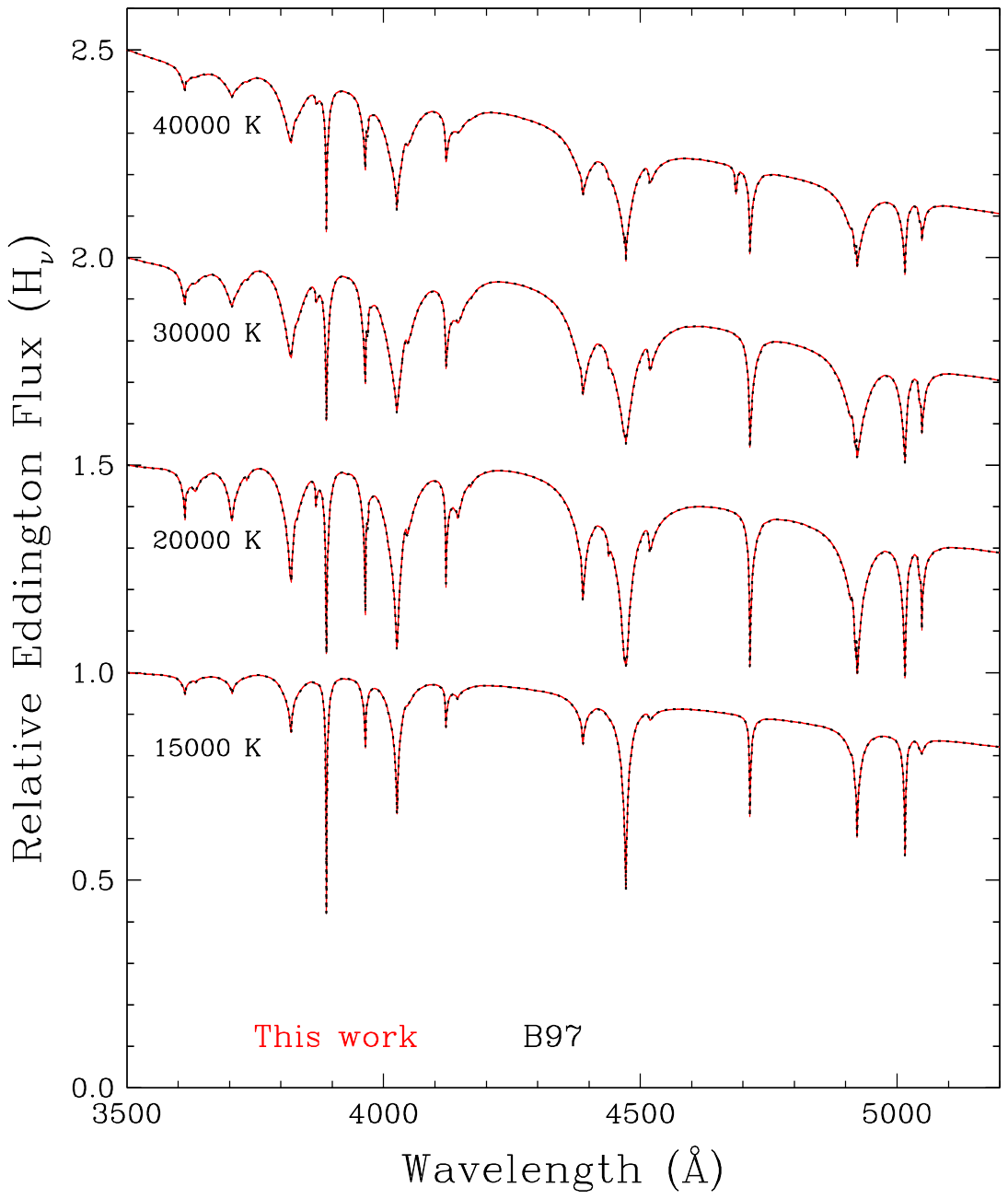}
\caption{Comparison of pure helium synthetic spectra in the optical
  wavelengths calculated using He {\sc i} Stark profiles from
  this work (solid red lines) and \citet[][B97; dotted black lines]{Beauchamp97} 
  at $\logg=8$ and various effective temperatures
  indicated in the figure. For the purpose of this comparison, line
  dissolution has not been included in any of these calculations.}
\label{fig:synth1}
\end{figure}

Figure~\ref{fig:synth1} presents a comparison of synthetic spectra
computed using the LTE model atmosphere code of \citet{Bergeron2011}
for pure helium compositions at $\logg=8$ and effective
temperatures spanning the range characteristic of helium-line (DB)
white dwarfs in which Stark broadening dominates
($15,000\lesssim\Te\lesssim40,000$~K). The spectra were calculated
using both the B97 line profiles and our improved Stark profiles
derived from computer simulations; results obtained at other values of
$\logg$ are qualitatively similar. It should be noted that the
semi-analytical profiles adopted here are not identical to those
originally published in B97, but incorporate several refinements
motivated by comparisons with our simulation-based profiles (see
Section~\ref{subsec:norm}). Since line dissolution is not included in
our simulations, it has also been omitted from the B97 profiles to
ensure a consistent comparison.

As expected, at the typical electron densities found in DB white dwarf
photospheres ($N_e \sim 10^{17}$~cm$^{-3}$), the semi-analytical B97
profiles and the simulated profiles are nearly indistinguishable. In
this regime, the effects of ion dynamics are therefore
negligible. This close agreement not only confirms the robustness of
the B97 calculations, but also provides an important validation of our
numerical simulations. In the context of white dwarf atmospheres,
however, the inclusion of line dissolution remains a crucial
ingredient. While it can be readily incorporated into the
semi-analytical profiles (see equations 4.1 and 5.5 of B97), it is not
yet implemented within our current simulation framework.

\begin{figure}[t]
\centering
\includegraphics[clip=true,trim=1cm 7.0cm 1.0cm 7.5cm,width=\columnwidth]{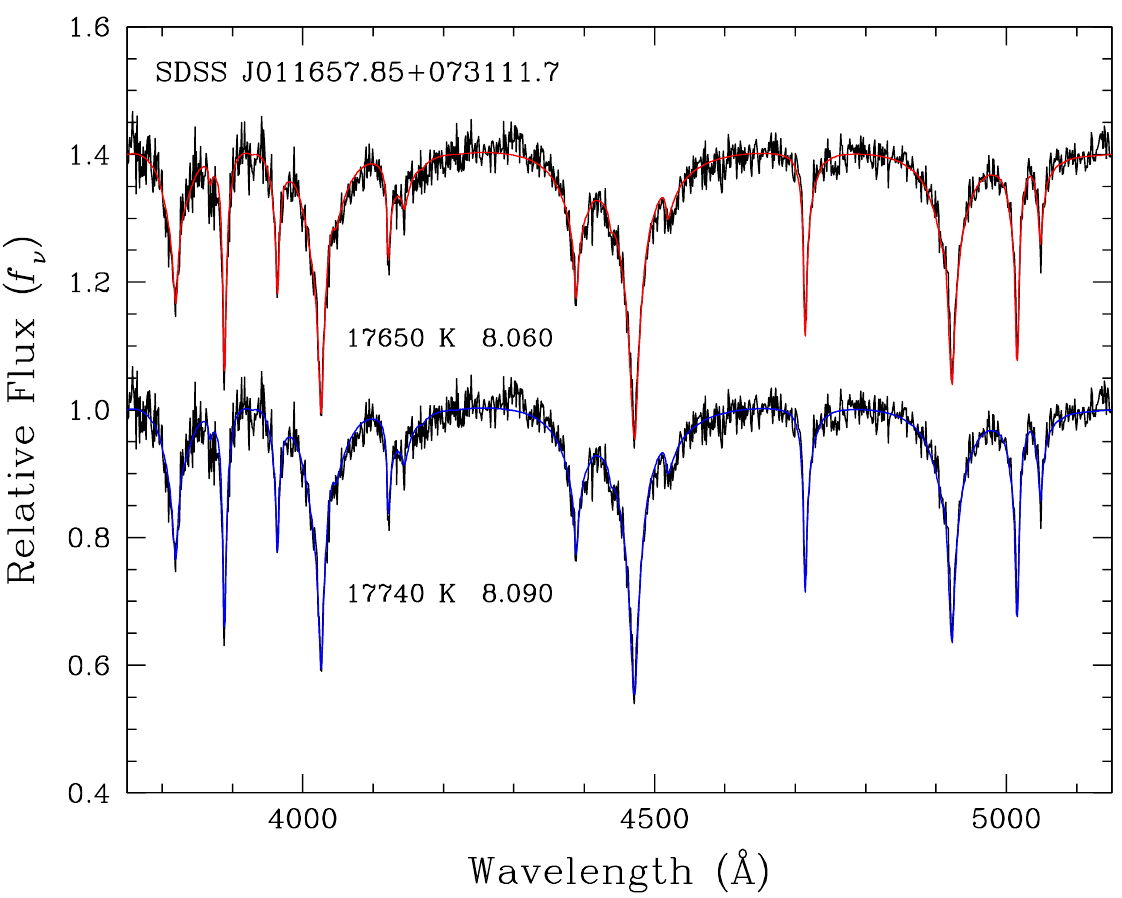}
\caption{Best fits to a representative DB white dwarf in the Data
  Release 17 of SDSS using pure-helium model spectra calculated with
  the He {\sc i} Stark profiles from B97 (red) and this work (blue);
  line dissolution is neglected in both cases. The $\Te$ and $\logg$
  values for each fit are given in the figure.}
\label{fig:compfit}
\end{figure}

For completeness, Figure~\ref{fig:compfit} shows the best fits
obtained for a representative DB white dwarf
(SDSS~J011657.85+073111.7) from the SDSS Data Release~17 (DR17), using
model spectra computed with both the B97 and the simulation-based
He~{\sc i} profiles. The resulting fits are qualitatively very
similar, and the inferred atmospheric parameters differ only
marginally, with the simulation-based profiles yielding an effective
temperature higher by $\sim$100~K and a surface gravity larger by
0.03~dex. A comprehensive analysis of the full sample of DB white
dwarfs identified in SDSS~DR17, based on both sets of line profiles,
will be presented in a forthcoming paper.

\subsection{Barnard 29} \label{subsec:bar}

\citet{Dixon2019} present a detailed analysis of several high-quality
spectra of the UV-bright star Barnard 29, located in the globular
cluster M13 (NGC 6205), including a very high-resolution spectrum
obtained with the Keck High Resolution Echelle Spectrometer
(HIRES). They derive atmospheric parameters of $\Te=21,400\pm400$~K,
$\logg=3.10\pm0.03$, and $\log N({\rm He})/N({\rm H})=-0.89$, by
comparing photospheric abundances of various elements and by fitting
the Balmer lines. In their analysis, NLTE model atmospheres are
computed using version 205 of the TLUSTY program \citep{TLUSTY}, while
synthetic spectra are generated with version 51 of the SYNSPEC program
\citep{SYNSPEC}. Their Figure~6 (see also their discussion in
Section~4.1.2) presents the best fit to the He~{\sc i} lines observed
in the optical HIRES spectrum of Barnard 29. They also discuss the
difficulty in reproducing the forbidden components in the blue wings
of He~{\sc i} $\lambda$4388 and $\lambda$4922. An improved fit to the
He~{\sc i} $\lambda$4922 feature (see their Fig.~7) is obtained by
replacing the line profiles from \citet{Shamey69} in SYNSPEC with
those from \citet{Barnard75}, which are computed for lower electron
densities of $10^{13}$~cm$^{-3}$, characteristic of the line-forming
region in Barnard 29. However, as noted by \citet{PTremblay2020}, the
line profiles from these two sets of calculations also differ in their
underlying physical assumptions. In particular, the profiles from
\citet{Barnard75}, based on the work of \citet{Barnard74}, incorporate
the one-electron approximation of \citet{Baranger62}, which provides a
more accurate transition between the impact approximation and the
quasi-static contribution of the electrons, as well as corrections for
ion dynamics.

\begin{figure}[t]
\centering
\includegraphics[clip=true,trim=2.2cm 2.5cm 2.2cm 3.8cm,width=\columnwidth]{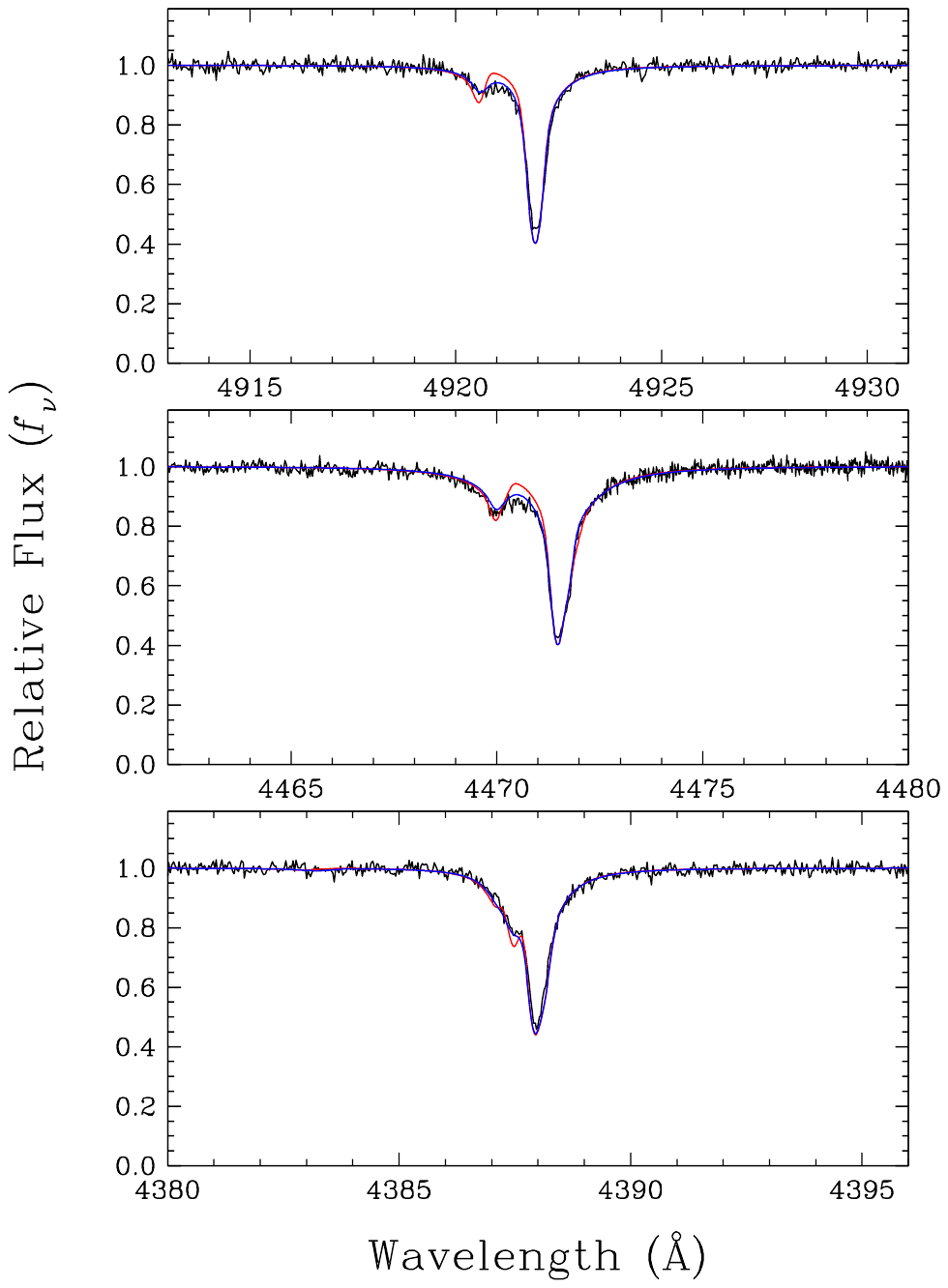}
\caption{Comparison of the optical spectrum of the post-AGB star
  Barnard 29 (black) with synthetic spectra calculated using the
  He~{\sc i} Stark profiles from \citet[][red]{Beauchamp97} and this
  work (blue), both including Doppler broadening. The synthetic
  spectra are convolved with a Gaussian of ${\rm FWHM}=0.1$ \AA\ to
  replicate the observed HIRES spectrum.}\label{fig:Barnard29}
\end{figure}

In this work, we reanalyze the HIRES spectrum of Barnard 29 (kindly
provided by P.~Chayer), using the same versions of TLUSTY and SYNSPEC,
into which we have incorporated our own sets of He~{\sc i} Stark
profiles (see \citealt{Bedard2020} for details). We adopt the same
atmospheric parameters as \citet{Dixon2019} given
above. Figure~\ref{fig:Barnard29} compares model spectra computed with
SYNSPEC using the He~{\sc i} Stark profiles from B97 and those derived
from our computer simulations for three neutral helium lines of
interest: He~{\sc i} $\lambda\lambda$4388, 4471, and 4922. For this
low-$\logg$ star, our improved line profiles provide a significant
improvement over the semi-analytical profiles for He~{\sc i}
$\lambda\lambda$4471 and 4922, owing to the explicit inclusion of ion
dynamics in our simulations. It is worth noting that \citet{Dixon2019}
already obtained a satisfactory fit for He~{\sc i} $\lambda$4471 (see
their Fig.~6); however, this result stems from their use of the line
profiles from \citet{Barnard74}, which already include partial
corrections for ion dynamics, as discussed above. By contrast, the
effects of ion dynamics on He~{\sc i} $\lambda$4388 (bottom panel of
Fig.~\ref{fig:Barnard29}) are less pronounced. We also note that the
weak forbidden component of this line is reasonably well reproduced by
both sets of calculations, in contrast to the results of Dixon et al.,
where this feature is not predicted (see the top panel of their
Fig.~6).

Two distinct improvements are identified, consistent with the effects
produced by ion dynamics, as discussed in
Section~\ref{sec:4471}. These effects have long been recognized for
the $2P^3-4D^3$ $\lambda$4471 transition, where the gap between the
allowed component and the $2P^3-4F^3$ $\lambda$4470 forbidden
component is partially filled. A similar behavior is observed for the
$2P^1-4D^1$ $\lambda$4922 transition, together with its associated
forbidden components $2P^1-4F^1$ $\lambda$4921. The He~{\sc i}
$\lambda$4388 line exhibits more subtle differences, primarily
resulting in a smoothing of the blend formed by the $2^1P - 4^1F$ and
$2^1P - 4^1G$ forbidden components located approximately 0.5 \AA\ from
the line core.

Since the dominant charged perturbers in this atmosphere are protons
rather than singly ionized helium (He~{\sc ii}), the contribution of
ion dynamics is expected to be even more significant than in our
current models. It would therefore be highly desirable to construct,
in the future, a grid of Stark profiles similar to ours but explicitly
including proton perturbers. The two upper panels of
Figure~\ref{fig:Barnard29} are consistent with a stronger contribution
from ion dynamics, which would further fill the gap between the
allowed and forbidden components and yield an even closer agreement
with the observed spectrum.

\subsection{HD~144941}\label{subsec:hd}

The second object investigated in this study is the extreme
helium-strong star HD~144941, analyzed by \citet{Przybilla2021}, who
derived $\Te=22,000$~K, $\logg=4.20$, and a surface helium fraction of
0.950 by number, or $N({\rm He})/N({\rm H})\sim20$. Their quantitative
spectroscopic analysis employed a hybrid NLTE approach in which the
thermodynamic structure of the atmosphere was computed under the
assumption of LTE, while the emergent spectra were obtained from
separate NLTE line-formation calculations. In the present work, we
adopt the same full NLTE approach as for Barnard 29, and rely on our
modified versions of TLUSTY and SYNSPEC to compute model atmospheres
and synthetic spectra using the atmospheric parameters determined by
Przybilla et al.

\begin{figure}[t]
\centering
\includegraphics[clip=true,trim=2.2cm 1.0cm 1.5cm 2.0cm,width=\columnwidth]{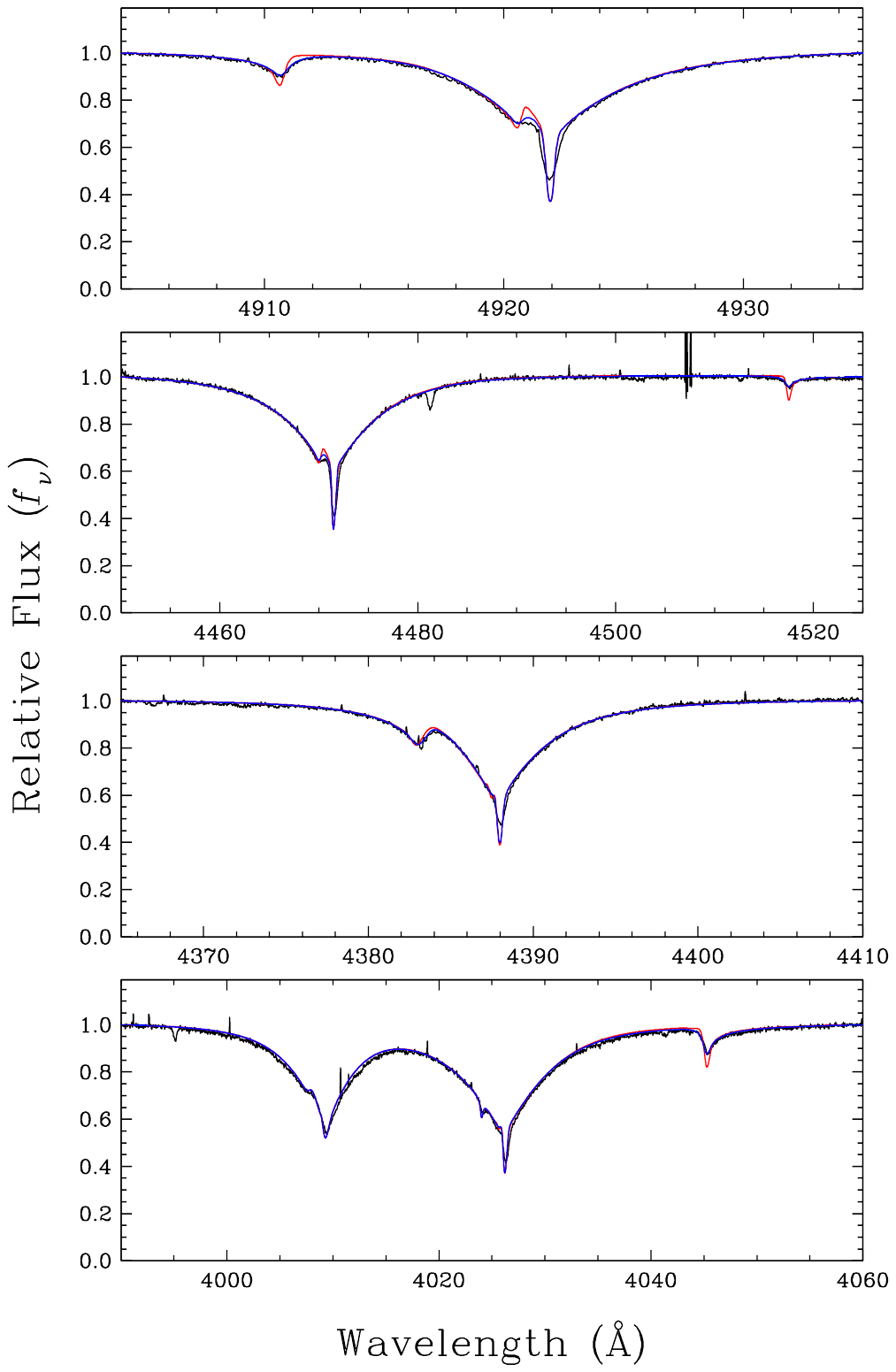}
\caption{Comparison of the optical spectrum of the extreme
  helium-strong star HD~144941 (black) with synthetic spectra
  calculated using the He~{\sc i} Stark profiles from
  \citet[][red]{Beauchamp97} and this work (blue), both including
  Doppler broadening. The synthetic spectra are convolved with a
  Gaussian of ${\rm FWHM}=0.1$ \AA\ to replicate the observed FEROS
  spectrum.}\label{fig:HD144941}
\end{figure}

We reanalyzed the optical spectrum of HD~144941 obtained with the
Fibre-fed Optical Echelle Spectrograph (FEROS) mounted on the
ESO–Max-Planck-Gesellschaft (MPG) 2.2\,m telescope at La Silla
(Chile), kindly provided by N.~Przybilla. Figure~\ref{fig:HD144941}
compares our best fits obtained with model spectra computed using the
He~{\sc i} Stark profiles from B97 and those derived from our computer
simulations, focusing on spectral regions around four neutral helium
lines of particular interest: He~{\sc i} $\lambda\lambda$4026, 4388,
4471, and 4922. In all cases, the profiles obtained from our
simulations yield a noticeably better reproduction of the observed
line shapes than the semi-analytical predictions, primarily as a
result of the more realistic treatment of ion dynamics.

Because this star has a higher surface gravity than Barnard~29,
additional forbidden components located farther from the line core
become apparent. The additional discrepancies observed when comparing
our newly computed profiles with those predicted by the semi-analytical 
approach are directly associated with these features. For the
$2P^3-4D^3$ $\lambda$4471 transition, the $2P^3-4P^3$ $\lambda$4517
forbidden component exhibits a less abrupt left edge. A similar
behavior is found for the $2P^1-4D^1$ $\lambda$4922 transition, where
the second associated forbidden component, $2P^1-4P^1$ $\lambda$4911,
also displays a smoother profile. To our knowledge, this is the first
time that such behavior has been successfully modeled for the
$\lambda$4026 and $\lambda$4388 lines as well. In the case of
$\lambda$4026, the forbidden component $2P^3-5P^3$ $\lambda$4045 shows
a noticeably smoother left slope compared to the prediction of the
semi-analytical theory. Likewise, the forbidden component $2P^1-4P^1$
$\lambda$4382 is more accurately reproduced, including the shape of
its right edge.

We note that the $2P^1-7D^1$ $\lambda$4009 line and the barely
detectable $2P^1-7S^1$ $\lambda$4024 line remain unchanged, as these
transitions were not included in the present simulation-based
modeling.

\section{Conclusion}\label{sec:conc}

In this work, we presented a new generation of Stark-broadened line
profiles for 13 neutral helium transitions in the optical range,
computed using modern computer-simulation techniques. By incorporating
an improved particle reinjection scheme and an updated power-spectrum
methodology, these calculations significantly extend the accuracy and
physical realism of earlier simulation-based efforts. The resulting
profiles exhibit more realistic wings at high density and an improved
reproduction of forbidden components—features that collectively reduce
several longstanding discrepancies when compared with the 
semi-analytical theory, as implemented by B97.

A detailed comparison with semi-analytical profiles based on the
standard Stark broadening theory reveals that the simulations
partially fill the gap between allowed and forbidden components at
lower densities and modify the shape of several isolated forbidden
components. These effects were previously documented for He {\sc i}
$\lambda 4471$ and $\lambda 4922$, but are modelled for the first time
in this work for additional helium transitions. These differences
directly impact spectral synthesis in astrophysical
applications. Although the effects remain moderate in white dwarf
atmospheres, they become more pronounced in lower-gravity objects,
such as Barnard~29 and HD~144941, for which ion dynamics plays a
larger role.

Despite these advances, the present simulations still lack the
inclusion of line dissolution and strong-collision quantum
corrections, and their accuracy may degrade at the highest densities
investigated. Potential further improvements include the incorporation
of multipole interactions \citep{Gomez2016}, the implementation of
full Coulomb simulations \citep{Stambulchik2022}, and the extension of
the quantum-state basis to enable a more accurate computation of
perturbed quantum states and their associated energies
\citep{Stambulchik2013,Cho2022}.

This study also highlights the challenges associated with generating
narrow line profiles—primarily those occurring at low densities—using
computer simulations. In this regime, hardware advancements are
required to obtain accurate and statistically reliable line
profiles. Alternatively, methods such as the frequency-fluctuation
method \citep[FFM,][]{Gigosos2014,Ferri2014} might offer a viable
solution for the lower-density parameter ranges. Nonetheless, the
profiles presented here represent a substantial step toward more
physically complete helium line grids. They also provide valuable
guidance for further improvements of semi-analytical models, several
of which were already implemented in this paper and are further
discussed in a forthcoming paper.

Overall, the simulation grid developed here establishes a modern
foundation for helium line-broadening calculations and offers a robust
framework for future spectroscopic analyses of helium-rich stars. Its
integration into atmospheric modeling codes will allow a more reliable
evaluation of Stark broadening effects and help refine our
understanding of the physical conditions in dense astrophysical
plasmas.

\begin{acknowledgements}
This work was supported in part by the NSERC Canada
and by the Fund FRQNT (Qu\'ebec).

\end{acknowledgements}

\section*{Data availability}
Both the semi-analytical (with and without line dissolution) and
simulation He {\sc i} line profiles are available on Zenodo, which can
be accessed via the following link
\dataset[doi:10.5281/zenodo.18722143]{https://doi.org/10.5281/zenodo.18722143}.

\bibliographystyle{aasjournal}
\bibliography{ms}

\end{document}